**Title**: Transformation of social relationships in COVID-19 America: Remote communication may amplify political echo chambers

**Authors**: Byungkyu Lee[1]*, Kangsan Lee[2], Benjamin Hartmann[3]

**Affiliations**:
[1]Department of Sociology, New York University; New York, USA.
[2]Social Research and Public Policy, New York University – Abu Dhabi; Abu Dhabi, UAE.
[3]Department of Sociology, Indiana University; Bloomington, USA.
*Corresponding author. Email: bklee@nyu.edu.

**Abstract**:
The COVID-19 pandemic, with millions of Americans compelled to stay home and work remotely, presented an opportunity to explore the dynamics of social relationships in a predominantly remote world. Using the 1972-2022 General Social Surveys, we found that the pandemic significantly disrupted the patterns of social gatherings with family, friends, and neighbors, but only momentarily. Drawing from the nationwide ego-network surveys of 41,033 Americans from 2020 to 2022, we found that the size and composition of core networks remained stable, though political homophily increased among non-kin relationships compared to previous surveys between 1985 and 2016. Critically, heightened remote communication during the initial phase of the pandemic was associated with increased interaction with the same partisans, though political homophily decreased during the later phase of the pandemic when in-person contacts increased. These results underscore the crucial role of social institutions and social gatherings in promoting spontaneous encounters with diverse political backgrounds.

**Teaser:**
COVID-19 showcased the resilience of core human social networks but also an alarming rise in political echo chambers.

## Introduction

Crises are generally observed to bring people together, deepen social ties, and strengthen communities, and it is a function of these processes that helps individuals navigate stress, hardship, and uncertainty that follow in the wake of disasters (*1*, *2*). However, the COVID-19 pandemic was different from other natural disasters as emergent norms and state regulations enforced "social distancing" that hampered face-to-face interactions and social gatherings necessary for maintaining social ties (*3*). One of the most profound shifts brought by the pandemic was the movement from in-person communication to a world where remote communication became a lifeline to our social lives. Unlike in-person gatherings across various interactional foci that foster spontaneous encounters (*4*), remote communication requires individuals to curate their social interactions more deliberately. This paper aims to understand how this transition to remote communication, with its inherent selectivity, affected personal network dynamics during the COVID-19 pandemic.

Before the COVID-19 era, academic discourses have centered around the rise of social isolation and the decline of social capital in America. In the early twenty-first century, Robert Putnam posited a decline in social capital due to television and the internet disrupting traditional community and kinship ties (*5*). A substantial piece of evidence comes from McPherson, Smith-Lovin, and Brashears' (*6*) analysis of the General Social Surveys, showing a substantial decrease in the size of Americans' core discussion networks over two decades from 1985 to 2004. However, subsequent research has suggested that this evidence may be influenced by methodological artifacts (*7*–*9*). On the other hand, research shows that a sense of social isolation could arise from the politicization of topics considered "important matters" under heightened political polarization (*10*) through the process of political echo chambers within core relationships (*11*). During the 2016 US presidential election, Americans discussed "important matters" with a small number of confidants who share similar political views (*12*) and cut their close relationships with politically dissimilar others (*13*). This trend raises concerns, as traditional offline networks were believed to nurture political disagreement, essential for fostering a democratic society, even amidst the growth of online echo chambers (*14*).

The presence of diversity in our relational environment, even in the face of a notable preference for homophily, is foundational to the fabric of society (*15*). Social ties are formed and maintained on the basis of individual preferences but only when suitable structural opportunities arise (*16*). These opportunities are traditionally constrained by physical boundaries, limiting the full potential to connect with like-minded individuals who are not in close proximity, thereby adding the element of randomness into social interactions (*17*). However, the advent of the Internet and remote communication has shifted this paradigm; Wellman and his colleagues argue that the Internet and new media facilitate social interaction beyond physical boundaries, bringing us to a networked society (*18*, *19*). While some academics have expressed concerns that new digital technologies may contribute to social isolation (*5*), empirical evidence suggests that individuals employ both face-to-face interactions and remote channels to maintain and further expand their social connections (*20*). Nonetheless, it is difficult to evaluate the extent to which network connectivity and homophily are shaped by physical boundaries and/or remote channels, since they are also shaped by individual preferences. In this regard, the COVID-19 pandemic presents a unique opportunity to study the dynamics of social relationships in a predominantly remote world, given that the pandemic has led to millions of Americans staying at home and working remotely.

The unprecedented impacts of the COVID-19 pandemic on virtually all aspects of American society serve as a test bed for the resilience of core relationships. Using various social metrics spanning four decades, Claude Fischer demonstrated that the resilience of core relationships has remained relatively unchanged since the 1970s, while peripheral relationships were more prone to fluctuations (*21*). Even when shifts appear to occur within core relationships, it may not be the relationships themselves that are transforming but rather the ways in which we maintain them that are evolving. However, it is plausible that the lack of substantial changes in core relationship patterns may be attributed to the absence of dramatic social changes that could influence them. Against this background, the lockdowns, a ubiquitous response to the COVID-19 pandemic, have disrupted traditional organizational focal points, reducing face-to-face interactions predicated on these foundations. It is crucial to examine whether the fabric of our social relationships, in their structure and dynamics, has been resilient enough to weather the profound transformations ushered in by the pandemic.

How would social relationships change during the COVID-19 pandemic that disrupted interactional foci such as in workplaces, voluntary organizations, and neighborhoods? Social distancing pressure and fear of infections likely reduced opportunities for physical contact and social gathering, which may have led to the thwarting of weak ties that could otherwise be enabled through serendipitous face-to-face interactions and participation in community activities (*5*, *6*, *22*). However, it does not necessarily imply that core networks would become smaller. With individuals increasingly using digital communication for social interaction (*23*), they may maintain their contacts and even form new relationships through these channels, which were increasingly available during COVID-19, leading to an increase in network size (*24*).

The decreases in the opportunity for spontaneous encounters, along with the widespread adoption of remote communications, might have strengthened the role of individual preferences in network dynamics, thereby replacing difficult relationships with easier ones. According to the theory of tie activation, when deciding with whom to discuss their important matters, individuals may deliberately mobilize social ties that they prefer or spontaneously use those that are readily available at the moment (*25*–*27*). The use of remote channels, in this regard, may enhance the role of the deliberative process, leading to an increased level of political homophily within core relationships. Simultaneously, the disruption of interactional foci would have decreased exposure to non-kin weak ties that are likely more politically heterophilous, thereby reducing the role of the spontaneous process promoting political diversity. This trend is more likely to occur during the pandemic as institutional constraints have been disrupted, potentially affording individuals greater leeway to sidestep challenging and onerous interactions with non-relatives, in contrast to the unavoidable engagements with kins dictated by familial duties (*28*).

To examine the multifaceted influence of COVID-19 on the patterns of social relationships, we relied on two main data sources. First, we conducted three nationwide ego-centric network surveys (i.e., COVID-19 network study) – the first one from April 2020 to April 2021, the second one in November 2021, the COVID Delta era, and the third one in May 2022, the COVID Omicron era. The unique strength of our survey lies in the use of the "important matters" name generator that identifies close confidants with whom people discuss "important matters," which had been widely used to characterize core discussion networks in eight national surveys from 1985 to 2016. These networks are known to include not only close families and friends but also those who are knowledgeable about important matters and those available when they arise (*26*), representing an important interpersonal environment for the exchange of information, influence, and social support (*6*, *29*). Second, we use the repeated cross-section General Social Surveys (GSS) spanning from 1972 to 2022. This dataset enabled us to compare the trends of social gatherings with families, friends, neighbors, and others at the bar before-COVID-19 era in a consistent manner (see Materials and Methods for details).

Exploring the changes in social relationship patterns during the pandemic presents us with a number of challenges to overcome. First, it is crucial to discern whether the observed changes in social relationships during the pandemic are a result of pre-existing trends or unique to the pandemic itself. To estimate the expected trends, we employ a multilevel meta-analysis that utilizes data from all available national ego-centric network surveys conducted prior to the pandemic. Based on the results, we establish a benchmark for comparison — either the identified trends, if any, or the overall mean in their absence. We apply the same approach to create a benchmark for patterns of social gathering, using all pre-pandemic data points in the GSS from 1972 to 2018. In both analyses, we incorporate data from the later stages of the pandemic, which help us assess the pandemic's lasting impact and determine if patterns eventually revert to pre-pandemic levels or persist. Second, we address potential issues of comparing results from the previous nationally representative, probability samples utilized in prior studies with the non-probabilistic nature of our COVID-19 network study. Specifically, we filtered out poor-quality responses and conducted a raking procedure to create post-stratification weights. The resulting sample weights allowed us to estimate weekly vaccination rates and race-specific COVID-19 infection rates, closely tracking those from the CDC (fig. S2 and fig. S3) from April 2020 to March 2021. Finally, we addressed the contextual sensitivity of the "important matters" name generator (*29–31*) (i.e., social contexts may shape what people consider to be "important" which in turn influences whom they talk to) by employing multiple name generators (important, health, and political matters). For a direct comparison with prior ego-centric network data, we primarily present results from the "important matters" name generator, though our main results are consistent with those derived from the multiple name generators.

Even with these rigorous approaches, it is crucial to recognize that the patterns identified in our COVID-19 network study should be interpreted with caution. First, it is a well-acknowledged fact that the reported characteristics of a network are influenced by factors such as the mode of survey, survey designs, and the question wordings used in network name generators (*29*, *32*). Although concerns could be partially mitigated by using wider prediction intervals estimated from our benchmark data that include a diverse range of pre-pandemic datasets, it is important to note that our COVID-19 network data were collected through opt-in, non-probability panels conducted exclusively online. Second, while we have applied statistical methods to estimate linear trends as benchmarks, existing literature indicates that network patterns, particularly political homophily, can fluctuate in different contexts, influenced by political events occurring at various times within the same year, such as elections (*10*, *11*, *13*, *14*). However, considering non-linear trends might lead to overfitting due to the data limitation, and the benchmarks incorporating data collected at various moments would result in larger confidence intervals in case the assumption of a linear trend is not valid. These challenges, nonetheless, underscore the critical need for our study, as it illuminates the profound ways in which the pandemic has reshaped network dynamics in America.

## Results

### *Social gathering during the COVID-19*

The COVID-19 pandemic seemed to change how Americans interact with one another, but it remains unclear which types of social interactions changed and for how long. Fig. 1 shows trends of social gatherings in the United States from 1972 to 2022 by tracking the percentage of Americans who spent social evenings more often than once a month with relatives, friends, neighbors, and others at the bar from the GSS data. Here, we find a general upward trend in spending social evenings with relatives, no change in social gatherings with friends, and a downward trend for social gatherings with neighbors or others at the bar. Surprisingly, during the pandemic, all these trends were reversed: a substantively smaller proportion of people spent social evenings at least once a month with their relatives (19% decrease), friends (26% decrease), and neighbors (32% decrease) in 2021 compared to 2018. However, the social gathering patterns swiftly returned to the pre-pandemic level in 2022, which suggests that the pandemic may have shifted the patterns of social gatherings only momentarily. Interestingly, social gatherings with relatives in 2022 still appeared to be slightly below the trend line, unlike those with friends and neighbors. It would be partly because people might have been still cautious when getting together with their older family members who are more likely to be vulnerable to the COVID-19 infection. We obtain similar results if we transform the response category into approximate daily units (i.e., almost daily = 7 × 52, several times a week = 3.5 × 52, several times a month = 1× 52, once a month = 1 × 12, several times a year = 4, once a year = 1, never = 0) or use other categories. Our findings on the temporary reductions and then immediate recovery of the social gathering patterns are generally consistent with mobility patterns recovering from the pandemic influence (*33*). From these patterns alone, however, it is hard to confirm whether it indicates that Americans were more socially isolated during the pandemic because they instead could maintain their social connections remotely.

### *Network size and social isolation during COVID-19*

Next, we examine how the overall size of core discussion networks – in which people confide their “important matters” – changed over time. To ensure the validity of our findings and rule out the possibility of trending effects, we establish a benchmark on the size of core discussion networks from the meta-analysis that uses all available data from 1985 and 2016. Fig. 2 shows that the average network size of core discussion networks follows the declining trends from 1985 to 2022 (also see Table S1). While there has been ongoing debate regarding the authenticity of the network size reduction from 1985 to 2004 — whether it is a real decline or a result of methodological artifacts — it is notable that the core network sizes have never rebounded to match the size in 1985. Seemingly, Americans have engaged fewer people to discuss their “important matters,” but it could reflect changes in what people consider as “important matters” (*10*).

During the pandemic, Americans discussed important matters with an average of 1.68 confidants in the early COVID-19 pandemic from April 2020 to April 2021, which is similar to those later from both the COVID Delta (mean = 1.63) and COVID Omicron era (mean = 1.62). While the size of the core discussion network was smaller during the pandemic than before excluding the strikingly small network size (1.38) in 2016, it did not significantly deviate from the meta-analysis benchmark trends, which makes it hard to conclude that it is the pandemic that caused core discussion networks to decrease. Despite debates surrounding the 2004 GSS as a potential outlier and concerns over the web-based survey designs of the 2010 and 2016 TESS—especially given the alarmingly increased level of social isolation in the latter—supplementary analyses including or excluding these datasets consistently showed the robustness of our findings (fig. S4 and fig. S4). We have also noticed a similar pattern for isolation in core discussion networks; 13.3%, 15.6%, and 15.7% of people reported having no one to discuss important matters with across three phases of the pandemic respectively, which again do not significantly differ from the meta-analysis benchmark (fig. S6).

While the number of Americans reporting isolation is small, it will be a greater concern if most social isolation arises due to relational isolation ("they have no person to talk to") rather than a lack of interest ("they have no important issues to discuss"). For example, one may wonder that as COVID-19 and the 2020 US presidential election brought up many important health and political issues, it is unlikely that people would report that they have no important issues to discuss. Nevertheless, we found that about half of isolated cases reported a lack of interest (i.e., 47.9%, 41%, and 45.7% across three phases respectively), and about another half of reported relational isolation (45.7%, 48.2%, and 43.3% across three phases respectively). The level of relational isolation during COVID-19 is roughly similar to previous reports (e.g., 43.4% reported by Lee and Bearman (*12*), 36% by Brashears (*34*), and 44% by Bearman and Parigi (*30*)). These results together suggest that Americans were not more socially isolated during the COVID-19 pandemic than the expected trend.

***The nature of relationships during COVID-19***

During the COVID-19 pandemic, it is possible that Americans could turn to different confidants to discuss important issues while the size of their core networks stayed the same. They could have activated kin ties they trust, generally stronger than non-kin ties, to cope with the uncertainty and the risk of disease transmission (*22*). To explore this possibility, we compared the relationship composition in core networks during COVID-19 against those from eight ego-centric network surveys from 1985 to 2016. Fig. 3 shows the patterns of relationship compositions in core discussion networks over the past four decades (also see Table S2). In general, kin ties comprised approximately two-thirds of the core discussion networks, while non-kin ties accounted for the remaining third. Again, the composition of activated relationships in core discussion networks remained consistent over the past four decades, except for small downward trends in the composition of neighbor and co-worker ties.

During the initial phase of the pandemic, Americans mainly activated strong ties, such as a spouse (25.6%), parents (12.8%), children (11.7%), siblings (9.9%), and other family members (7.4%), rather than weak ties like friendship (23.2%), neighbors (1.6%), and coworkers (3.8%). These patterns were consistent in the later phases of the pandemic. Our survey estimates mostly align with the range of estimates from the meta-analysis, indicating that the COVID-19 pandemic did not substantially disturb the entrenched patterns of relationship composition in core discussion networks, barring two exceptions. Americans were slightly more likely to confide with their children or neighbors compared to the established trends, yet these deviations were minor and lost significance in the later stages of the COVID-19 pandemic.

***Network homophily during COVID-19***

The stability and resilience regarding the size and composition of Americans' core networks during the pandemic do not necessarily imply that their makeup has remained unchanged. The replacement of network ties may induce changes in network homophily due to the network churning process (*35*). As discussed, the proliferation of remote channels may allow individuals to activate ties with non-kins with similar political leanings and the disruption of interactional foci may decrease exposure to non-kin ties who are more politically heterogeneous. Here, we examine the level of homophily in these networks by measuring the extent to which our close confidants resemble us. Fig. 4 displays the levels of absolute homophily with 95% confidence intervals across five different characteristics from 1985 to 2022. Prior to 2020, Americans' core networks were characterized by increasing trends in educational homophily and decreasing trends in racial homophily across both kinship ties and non-kin ties. While there was no evident decreasing nor increasing trends in political homophily in the pre-COVID era, political homophily notably escalated during the pandemic, especially among non-kin ties. For example, in 1987, about 52% of confidants shared the same partisanship as respondents, which increased to 68% in 2020. However, the surge in political homophily appeared to lessen in the later phases of the COVID-19 pandemic.

Nevertheless, it is important to consider that these changes in absolute homophily might potentially reflect the changes in demographic compositions and ideological distributions in the U.S. or the changes in the distribution of network sizes (*36*). To address these issues, we identify choice homophily by employing random mixing models (see Materials and Methods). Fig. 5 shows the patterns of choice homophily across socio-demographic characteristics and partisanship, in which the dotted line at one represents the default result expected from random mixing, whereas values higher than one denote significant homophily. These analyses reaffirm the general principle of network homophily – "similarity breeds connection" (*37*) – with the exception of sex.

One interesting observation from the meta-analysis is that racial choice homophily has increased especially among kin ties. Our additional analysis of racial choice homophily by different racial groups (see fig. S7) reveals that this increase was larger among whites than among other racial groups (Panel C). Specifically, from 2008 to 2016, we observed a substantial increase in racial choice homophily among whites alongside an increase in the proportion of "other racial groups" in the United States (Panel A). This result is broadly consistent with existing work showing that whites seek to reinforce a white/Non-white divide following the growth in the Hispanic population (*38*), though it is surprising to observe a similar pattern in actual social relationships rather than hypothetical and stereotyped racial groups.

The most notable change during the COVID-19 pandemic is a sharp rise in political choice homophily, especially among non-kin ties. The general pattern – lower levels of political choice homophily among non-kin ties than among kin ties – was reversed during the pandemic such that political choice homophily was similar between non-kin and kin ties. By decomposing political choice homophily across different partisan groups (see fig. S8), we find that choice homophily among Republicans and Democrats is larger than those among Independents and non-voters, though the level of political choice homophily has increased to a similar extent across all four groups across both kin and non-kin ties since 2016. Notably, the COVID-19 pandemic has led to a particularly striking rise in political choice homophily among non-kin ties. This is concerning because non-kin ties, such as those with co-workers and friends, typically facilitate opportunities for cross-ideological interactions and political deliberation (*14, 39*). Finally, following the initial phases of the pandemic, there was a decline in political choice homophily in subsequent periods. This trend implies that certain factors unique to the pandemic might be driving the emergence of political echo chambers within personal networks, though these effects may be temporary.

Our findings so far demonstrate that Americans were more likely to rely on politically similar confidants during the pandemic without dramatic changes in network size or relationship composition. We now shift our attention to the role of remote communication to explore how Americans' pandemic responses using remote channels against social distancing pressure and fear of infections might influence the observed increase in political homophily within personal networks.

***The role of remote communication channels during the pandemic***

One major social change prompted by the pandemic was the rapid adoption of remote communication channels, such as Zoom, across various organizational and institutional contexts. To examine how people stayed connected with others despite physical distancing, we asked which communication channels people used in their recent conversations with each confidant. Fig. S9 shows that Americans, during the initial moments of COVID-19 until April 2021, activated 58.6% of their ties through in-person contacts, followed by phone (45.3%), text messages (34.7%), video calls (14.5%), social network services (9.5%), email (8.0%), and other channels (1.9%). This preference for traditional communication channels over new technologies is consistent with the earlier results from the 2008 PEW survey (*40*). As the threat of COVID-19 diminished due to a combination of reduced virus severity and widespread vaccinations, the proportion of those who use in-person contact gradually increased during the later stages of the pandemic (62.3% during the COVID Delta wave and 64.5% during the COVID Omicron wave), but the proportion of those who used video or social network services for core discussions declined. Given that core discussion networks often comprise family members who reside in the same residence, it is crucial to assess how much people rely on within-household or between-household ties throughout the pandemic. In our survey, approximately 37.9% of confidants lived in the same household during the initial pandemic period, which decreased gradually in later pandemic periods (36.8% during the COVID Delta wave, and 35.8% during the COVID Omicron wave). In contrast to the 22.2% documented in the 2008 PEW survey, our findings suggest that during the pandemic, Americans were more likely to mobilize in-home ties.

Next, we compare the distribution of network ties across geographic locations and communication channels between the 2008 PEW survey and our COVID-19 survey (see Table S3). During the early pandemic, Americans showed two diverging patterns – engaging in in-person contacts with their confidants living in the same household (32.7%) or using remote communication channels with their confidants living in a different household (40.1%). The proportion of remote communication with confidants living in a different household increased by 18.7 percentage points, whereas the proportion of in-person communication with confidants in the same household increased by 9.4 percentage points during the pandemic compared to 2008. For example, in 2008, 54.7% of Americans used in-person contact with someone living in a different household, but during the pandemic, only 23.9% used those contacts. Additionally, the increase in in-person contacts in the later phases of the pandemic mainly arose among different household relationships. So far, these results show that Americans managed to maintain close social relationships during the pandemic through face-to-face interactions with alters living in the same residence or to maintain connections with others who were geographically distant via remote communication channels.

What could be the implication of the diverging patterns of geographically proximate face-to-face interactions and geographically distant remote interactions for political homophily? To address this question, we conduct logistic regression models at the dyadic level to examine how the use of remote versus in-person communication channels is associated with political homophily while controlling for individual socio-demographic characteristics and alters' location. The results, shown in the left panel of Fig. 6, consistently replicate a general pattern wherein homophily is stronger among kin ties compared to non-kin ties (*41*). Specifically, the sole use of remote communication channels or a combination of both remote and in-person channels, as opposed to the sole use of in-person channels, was associated with an increase of more than four percentage points in political homophily among non-kin ties. This supports the idea that the elevation of political homophily within non-kin ties during the pandemic was driven by individuals' unbounded preferences extending beyond the local and physical boundaries. On the other hand, the use of remote channels was associated with a decrease in political homophily among kin ties. This suggests that the activation of kin ties may not be solely driven by individual preferences. In situations where tie activation is not entirely an individual choice due to family obligations and norms, remote communication channels can serve as a medium for engaging with other families who may hold differing views, without the need for inperson gatherings during the pandemic.

The right panel of Fig. 6 illustrates the shifts over time in the relationship between communication channels and political homophily across three phases of the pandemic. In general, there was little variation in the levels of political homophily among both kin and non-kin ties across various communication channels throughout all three phases. However, two exceptions were observed within non-kin ties when people use exclusively in-person or remote channels. We found that as the threats posed by the COVID-19 pandemic gradually diminished, Americans increasingly confided their "important matters" with non-kins who hold differing political views, either exclusively through in-person contacts or remote channels. This shift is notable for its substantial effect size, with a difference of over eight percentage points compared to the early pandemic phase. This emphasizes the pivotal role of interaction foci in fostering spontaneous encounters with diverse political views and highlights the constraining nature of remote communication in reinforcing homogeneous political preferences. In other words, risks associated with in-person contacts might lead to a transition to remote channels, but this adaptation appears to have been accompanied by a tendency to engage more frequently with those who were politically similar.

Taken together, these results suggest that the pandemic situations that amplified social interactions through remote channels might primarily facilitate the process of tie activation driven by homophily only when tie activation was subject to individual choices, as seen in the case of non-kin ties. Differently put, the COVID-19 pandemic revealed the importance of social institutions that have exposed us to diverse people whom otherwise we might not have interacted with.

**Discussion**

The COVID-19 pandemic has presented a social dilemma; “social distancing” was necessary to curb the spread of disease, yet social connections were needed more than ever to collectively overcome the unprecedented crisis. Despite the strong push against in-person contacts, we discover that the size and relationship composition of core discussion networks did not diminish during the pandemic. In the face of crisis, individuals adapted the ways they maintain their relationships, from face-to-face interactions to remote interactions, demonstrating the resilient nature of core relationships. While it might not be surprising to some audiences that our core discussion networks are resilient and stable (*21*), it is still remarkable that network stability could be maintained against one of the most profound social changes regarding social interactions brought about by the COVID-19 pandemic.

Personal network dynamics literature shows that people activate and deactivate their social ties following different life events across individuals’ life courses, which is called “network churn” (*42*, *43*). At the micro level, social relationships must have changed during the transformative period, but how? Our results indicate that the COVID-19 pandemic has amplified political divides in core relationships, exacerbating the trends of rising interpersonal echo-chambers. We show that it is likely driven by the switching mode of communication; some social ties maintaining political diversity might be dropped due to the disruption of interactional foci, at least momentarily. The disruption of interactional foci caused by the pandemic has likely eliminated natural opportunities for spontaneous network activations. These foci are where people naturally interact with dissimilar others and sometimes discuss “important matters” that they may not be able to share with their close family and friends due to the fear of incompatible expectations (*25*). As people find it difficult to discuss “important matters,” including COVID-19, with others who may hold opposing views, the use of remote channels will facilitate the activation of homogeneous ties with those who are likely to share similar perspectives.

However, the rise of political homophily during the pandemic cannot be attributed solely to the mechanism of the use of remote channels and the disruption of foci. First, political homophily in our personal relationships could arise due to the combination of polarizing events and politicization processes. It is essential to recognize that the COVID-19 pandemic emerged not in isolation but against a backdrop filled with other large-scale polarizing events, including Black Lives Matter and Capitol Riot. In addition, increasingly partisan elections frame “important matters” as “political matters,” and such a framing was likely amplified during the 2020 election because the response to the pandemic immediately became highly politicized and politically divisive (*44*). Earlier works show that individuals are more inclined to activate politically similar ties, reduce family time during holidays such as Thanksgiving, and even disengage from politically dissimilar friendships in politicized situations such as contested elections (*11*, *13*, *45*). In our future work, we will investigate how polarizing events and politicization of the pandemic responses shape racial and political homophily.

In the aftermath of natural disasters, social cohesion frequently emerges as individuals facing similar challenges and concerns could connect with those who were previously unconnected, and they would be willing to help one another during community rebuilding processes (*46*, *47*). Yet, solidarity-inducing processes generally occur in tandem with processes that seek to blame others – we all know that scapegoating almost always wins out (*48*). Facing external threats increases trust and cooperation within in-groups but reinforces the boundary between "us" and "them," resulting in greater division between groups (*49*, *50*). The boundary-making process induced by exposure to common enemies is more pronounced in more polarized contexts (*51*). Within this framework, the demarcation of boundaries during a crisis, coupled with the politicization of pandemic responses, may contribute to Americans' increased likelihood of relying on the same partisan confidants who may have even extreme political views (*52*). In addition, the emerging pandemic precarity and health inequality associated with race and partisanship (*53*, *54*) can be useful for explaining this pattern. For example, millions of Americans who had to face the deaths of friends or families during the pandemic may have deactivated existing ties or formed new social ties during the pandemic (*55*, *56*). In doing so, if certain partisan groups were more severely impacted by the pandemic, then the observed patterns of homophily might be influenced due to the shrinking size of such groups.

The surprising resilience of Americans' core networks in the face of escalating psychological distress and loneliness during COVID-19 (*57*) is surprising, given that anxiety, loneliness, and depression are commonly linked to social isolation and inadequate social support (*58*, *59*). It also contradicts the notion that Americans were "socially" as isolated as they were physically (*60*). Here, the finding that Americans could uphold their core relationships through remote channels during the pandemic may imply that their feelings of loneliness and isolation may have arisen from shifts in specific types of social interactions, such as the decrease in face-to-face interactions with neighbors and friends (*61*) and the decline of social gatherings as shown in Fig 1. These seemingly peripheral relationships that were available in core networks play a vital role in social support. Consequently, the absence of these connections may disrupt the social fabric, contributing to a unique form of loneliness (*22*, *58*, *61*). These results invite future work to examine the implication of different modes of communication contributing to individuals' well-being.

During the COVID-19 pandemic, social networks displayed a mix of adaptability and consistency over time. In the early stages, there was a noticeable increase in political homophily via remote channels. This trend continued into the later stages, even though political homophily through face-to-face interactions dropped when social lives began returning to "normal." This sustained shift raises concerns about the formation of echo chambers through remote communication. Simultaneously, the over-time decline in political homophily in face-to-face interactions underscores the importance of various interactional foci that promote political diversity in social connections (*4*). It is crucial to delve deeper into both short-term and long-term changes in these shifts in political homophily and the role of various communication channels in shaping social relationships. In our ongoing analysis, we break down the weekly trends during the pandemic, showing that the patterns of remote communication and in-person interactions are strongly correlated with the combination of the viral transmission of COVID-19 and public attention to the pandemic. In addition, we plan to run another nationwide ego-centric network survey to examine the long-term consequences of rising political homophily when the COVID-19 pandemic is "officially" over.

The increasing trend of racial choice homophily in American society is notably concerning, even though it is not exclusively tied to the COVID-19 pandemic. Smith et al (*36*) who compared the patterns of homophily between the 1985 and 2004 GSS failed to identify these patterns given that the rise began after 2004, and their point-to-point comparison approach is unable to capture the trend. As the United States becomes more racially diverse, particularly with increases in Hispanic and Asian populations and immigrants, the composition of close relationships would become more racially diverse, leading to a decrease in absolute political homophily. However, the upward trend in racial choice homophily indicates that Americans increasingly prefer being connected with those who share similar backgrounds and values, a preference possibly driven by a sense of safety and social support in more homogeneous environments (*62*). It is well known that individuals prioritize community building and solidarity at the expense of minority groups in the face of external threats (*47*, *63*). The disruption of interactional foci limiting exposure to weak ties may have led core networks to become a place where individuals share not only information but also a sense of co-ethnic identity by drawing a distinction between us and them. As rising racial choice homophily poses a serious challenge, especially in its potential to amplify social segregation and fragmentation, it is concerning to observe that these processes unfold within our core relationships.

Our findings on the patterns of ego networks have several implications for the global network patterns. The stability in network size and relationship composition suggests that it is unlikely that the property of our global network structure has changed as well. Instead, the growing trend of political homophily in ego networks, facilitated by the use of remote channels, may reflect increased fragmentation and division globally in Americans' social networks. It suggests that even if solidarity were to increase in the face of the COVID-19 disaster, it would be likely to occur only within distinct partisan ingroups. The ongoing trends of political sectarianism equating one's political positions with their moral quality (*64*) may have been exacerbated by increased political homophily during the pandemic as core relationships are more likely to be embedded in multiple relational contexts where people might be more willing to attend to dissenting views and learn from other perspectives (*14*, *65*). It is particularly concerning because the creation of echo chambers and the reinforcement of preexisting biases can result in the reinforcement of entrenched political and ideological positions, making it more difficult for individuals to consider alternative viewpoints and contributing to further polarization.

Several limitations are worth noting. First, our online nationwide survey across 50 states and Washington D.C. is not a probability-based sample. There are reservations among researchers about the use of large surveys that may not be fully representative, as they can amplify survey biases (*66*). To address this issue, we carefully filtered out poor responses and used statistical techniques to account for potential sampling biases, which enabled us to closely track the official vaccination rates over time and the COVID-19 positivity rates from the CDC. Although we believe these benchmark results enhance the credibility of the estimates of network characteristics obtained from our surveys, we also acknowledge no single data source or benchmark can be entirely without limitations or potential biases. Therefore, even if the patterns from our surveys align well with these benchmarks, it would not conclusively establish our survey as "nationally representative."

Second, our strategy to deal with fraudulent responses may be more likely to erroneously exclude specific groups, for example, Republicans and racial/ethnic minorities. Although it is crucial for us to exclude problematic survey responses using established protocols to improve the quality of our survey, we acknowledge that doing so could introduce potential bias into our results. Furthermore, using an online opt-in non-probability sampling strategy could lead to increased selection bias, as respondents could have self-selected to participate in the survey for systematic reasons like interest in COVID-19 and/or aversion to politics. While these potential sources of bias are less likely to be found in previous studies that used probability-based sampling methods, it is crucial to note that even nationally representative surveys like the General Social Survey suffer from the low response rate (i.e., 17% in 2021) during the pandemic period.

Third, the absence of panel data on networks restricts our ability to directly identify network churn processes. Instead, our insights on churn are inferred by comparing observed networks during the pandemic to earlier estimates. As such, the nuances of adding new ties, maintaining existing ones, or dropping ties might not be captured comprehensively. Specifically, the aspect of dropped ties, which plays a vital role in understanding network churn, is not thoroughly explored in our current analysis. Finally, since all of our network data are based on an ego's self-reports on their alter, the increase in political homophily may reflect changes in perception rather than changes in reality. For example, it is conceivable that Americans were exposed to more political disagreement in their close social environments than this name generator would capture (*67*). Still, the increase in perceived political homophily would present considerable challenges to American society given that the perceived (potentially biased) network characteristics strongly shape individual attitudes and behaviors (*29*).

Despite these limitations, our findings on the structure of interpersonal networks during the COVID-19 pandemic have multiple implications. The phenomenon of increasing political homophily could create echo chambers that stifle democratic deliberation and lead to polarized discussions while simultaneously eroding social cohesion and trust. This fragmentation might even extend to health-related behaviors, with political alignments shaping attitudes towards infectious diseases, vaccine behaviors, mask-wearing, and trust in medicine (*68*). Such a trend may permeate beyond politics, affecting community engagement, information sharing, and public policy reception. In addition, remote channels have proven effective in sustaining social relationships during physical distancing (also see *69*), though the rising trends of political polarization through remote channels are concerning. As society increasingly gravitates toward online interactions, including remote work and virtual connections, these insights underscore the importance of fostering diverse interactions to maintain social cohesion. The pandemic's insights offer a nuanced perspective on the evolving nature of social connections in an increasingly virtual world, highlighting both opportunities and challenges. Our findings underscore the need for continued research to fully grasp the enduring effects of the COVID-19 pandemic on the polarization and cohesion of American society in the years ahead.

## Materials and Methods

## Data collection and quality control

Our COVID-19 network study consists of three nationwide ego-centric network surveys; the first survey collected approximately a hundred Americans each day from April 2020 to April 2021 (total N = 36,345), the second survey collected 1,776 Americans in November 2021, and the third one collected 2,912 Americans in May 2022, across 2,502 counties across 51 states. We recruited survey respondents from the Lucid Marketplace, which is made up of hundreds of suppliers with a diverse set of recruitment and sourcing methodologies (*70*). Since not all survey respondents who saw our survey in the Marketplace would complete it, we distributed our surveys three times a day (morning, afternoon, evening) to ensure that we had the similar number of completed responses every day. Informed consent was obtained from the survey participants before they took our survey. Fig S10 shows that we were able to collect about 100 responses per day on average (daily mean = 101.8, SD = 38.2) in our first survey, though there were some day-to-day fluctuations in sample sizes. Since we do not know how many people were invited to participate in this survey, we are unable to calculate official survey response rates. Prior to participation, all subjects provided informed consent. The study protocol received an exemption from review by the Institutional Review Board (IRB) at both New York University – Abu Dhabi and Indiana University.

Several recent reports have documented notable declines in the response quality from online panel surveys, such as Amazon MTurk and Lucid Market place during the COVID-19 pandemic (*71*, *72*). Furthermore, studies reported that many online panels did not pass attention checks, which could undermine the validity of survey responses (*73*, *74*). To address these concerns, we took the following measures and exclude fraudulent responses based on the recommendation by Kennedy et al (*74*). First, we carefully examined three open-ended text responses in our survey to identify nonsensical responses. These include (i) "name" fields in three network name generators (i.e., respondents are asked to write down either nick names, initials, or first names), (ii) fields explaining why respondents did not have someone to discuss important matters, political matters, or health matters, (iii) an open-ended question about past and current occupation. The first author initially reviewed all fields, which were then independently reviewed by two other authors. Then any discrepancies were reviewed again by all coders. As a result, we dropped 5,302 responses following this criterion (i.e., 3,904, 827, and 2,282 respectively for (i), (ii), and (iii)) Second, we excluded non-US respondents by coding the location of respondents' IP addresses following the recent methodological literature (*74*). We used two different IP address service locators (ipdata: https://ipdata.co/ and iphub: https://iphub.info/) to code the location of respondents' IP addresses. We identified 1,833 non-US IP addresses and dropped them. Although it would be possible for some respondents to take our survey through a proxy IP address using VPN (virtual private network), our examination of cases using VPNs showed that they had notably poorer knowledge on COVID-19 (see fig. S11). Third, we used ipdata (https://ipdata.com/)'s threat intelligence service to detect malicious IPs like malware sources, spam sources, botnets and blocked traffic from all IP addresses listed in any of 400+ blocklists with 600M bad IPs listed. We dropped 520 responses using this approach. Finally, we excluded those who completed the entire survey in less than five minutes since it was nearly impossible to finish this survey within such a short time (c.f., the median duration is 15 minutes). We conducted this review monthly throughout the entire fieldwork period and dropped 783 responses following this criterion. Fig. S12 shows the joint distribution of different types of fraudulent responses.

The grouped box plots in fig. S11 indicate that those who were classified as fraudulent respondents have poorer COVID-19 knowledge scores than our final sample. Table S5 shows the demographic characteristics of these bad responses, compared to our final sample. The fraudulent respondents were more likely to be male, young, non-white, Republicans, college graduates or higher education degree, married, working now or self-employed, living in Pacific or Mid and South Atlantic Division, and report either the lowest or the highest family income. Among 56,280 participants in our first survey, we found that 96.8% of respondents who clicked the survey link agree to participate in the survey, and 79.7% of those who agreed to participate completed the survey. Overall, these patterns are similar across three surveys, except that there was a decline of the fraudulent responses in the third phase. Fig. S13 describes the trends of survey participation status (those who did not agree to participate, dropped out, appeared to be fraudulent respondents, and the final analytic sample) over time. Using a correlational analysis of daily new COVID-19 case rates, we found that as COVID-19 cases rose nationally, individuals were less likely to decline participation ($r = -0.14$, $p < 0.01$). However, this effect on sample selection is offset by a rise in dropout rates ($r = 0.14$, $p < 0.01$), In conjunction with no discernible temporal patterns in the percentage of fraudulent respondents ($r = - 0.06$, $p = 0.19$), we found that the proportion of those included in the final analytic sample were not related to the COVID-19 transmission dynamics.

**Network name generators**

The ego-centric network survey consists of name generators that prompt respondents (“egos”) to think about their confidants (“alters”), and name interpreters to identify the characteristics of relationships and alters (*29*). We revised the General Social Survey’s classic instrument, the “important matters” name generator, to map core discussion networks (see Appendix B in SI for details). Critically, what people consider to be “important” shapes whom they talk to and thus invokes different kinds of confidants (*30, 34*). Before conducting our survey, we carried out a pretest, which showed that 61% and 41% of conversations within core discussion networks were about health and politics, respectively. We later confirmed that people discussed health and politics to the similar extent (health: 55%, politics: 43%) in our core discussion networks throughout the entire survey period. Consequently, we asked respondents to elicit up to five names from “important matters” name generators, and then asked up to three names from political and health matters name generators, respectively, in a randomized order, instead of using five question boxes based on the pretest (see Appendix A in SI for the details of our pretest). Then, we combine multiple name generators (important, health, and political matters) to address the contextual sensitivity of the “important matters” name generator (*31*). However, we primarily present results from the important matters name generator, although our conclusion stays the same if we use results from the multiple name generators. To ensure that we do not train our respondents to name fewer discussion partners, we locate our network questions at the beginning of the survey. Once we collect “names” (e.g., nick names, first names, initials) of each alter, we use name interpreters to collect information about them including the nature of relationship (e.g., relationship type, discussion topic and frequency, communication channel, the timing of last in-person contact, and geographic location of alters) as well as their demographic characteristics (e.g., age, gender, race, education, and partisanship). The exact question wording and response options are described in Appendix B in SI.

## Measures

*Network size, isolation, and other variables* We measure the size of core discussion networks by counting the names that appear in the important matters name generators. We identify network isolation when respondents do not provide any names. If respondents do not report at least one confidant in the important matters name generator, we further ask whether this is because they do not have anyone to discuss important matters with or because they do not have any important matters to discuss. For other variables collected by network name interpreters, we use the raw response categories unless otherwise noted, including relationship type and demographic characteristics of alters (i.e., age, gender, race, education, and partisanship).

*Relationship composition* When comparing the relationship composition of our survey to earlier surveys, we note that the earlier network surveys (GSS, CNES, PEW) allowed respondents to report multiple relationship categories for each alter, whereas later surveys (TESS, and COVID-19 study) only permitted reporting a single category that best captured the relationship. To address this discrepancy, we choose a relationship category from multiple ones by first randomly assigning a category from kin ties for alters who are tied to an ego through any kin relations, and then randomly assigning a category from non-kin ties. We repeat this procedure 1000 times to simulate the distribution of relationship composition in each survey.

*Homophily* We quantify absolute homophily and choice homophily. Absolute homophily is defined with respect to the similarity of an ego-alter pair, without considering the opportunity structure, whereas choice homophily is defined with respect to ego-alter similarities in reference to those expected from random mixing, conditional on the population composition and degree distribution (*37*). We measured absolute homophily by calculating the proportion of the same categorical attributes between ego and alters within each ego's network. To measure choice homophily, we extend the case-control matching strategy by simulating random mixing processes on ego networks (*36*). Specifically, we exploit the fact that individuals in a nationally representative survey data can be potential "alters" for each ego, who are unlikely to know each other. In doing so, we randomly select potential alters corresponding to the ego network sizes from all survey respondents excluding ego, and then generate 1000 simulated ego-centric networks. We then measure choice homophily by dividing the observed homophily by the mean chance homophily across 1000 simulations (*75*) . We also calculate Coleman index to capture choice homophily net of compositional differences across different groups using *netseg* package in **R** (https://github.com/mbojan/netseg/) (*76, 77*). To ensure consistency across different datasets, we recode demographic categories for ego and alter into the followings: age (less than 20, 20-39, 40-59, 60+), sex (male, female), race (white, Black, Other), education (less than high school, high school, college graduate and higher), and partisanship (democrat, republican, independent, something else).

## Analytic strategy

We used all available data including the GSS 1985, 1987, 2004, 2010, the CNES 1992, the PEW 2008, and the TESS 2010/2016 for comparison. While the information about network size and relationship type was available for all studies, other information was partially available across different surveys. Table S1 in SI summarizes the similarities and differences in study designs**.** To provides reliable benchmarks, we conducted a meta-analysis to summarize estimates on network characteristics prior to 2020. In doing so, we tested the significance of linear trends, and then we establish a comparison benchmark — either the identified trends, if any, or otherwise the overall mean from the following random effects models. This allowed us to assess whether the patterns of core discussion networks during the pandemic significantly differ from the general trends. Specifically, our random effects models capture the overall tendency as well as the linear trend if any, assuming that an estimate from each survey $y_t$ at period *t* is the combination of the unknown true effect ($\theta_t$) and the sampling error ($e_t$): $y_t = \theta_t + e_t$, where $\theta_t = \mu + \beta t + u_t$, $e_t \sim N(0, v_t^2)$ and $u_t \sim N(0, \tau^2)$. Namely, we assumed that differences in estimates across different surveys may introduce some random variability among the true effects in addition to the random sampling error. We used restricted maximum-likelihood estimation to estimate $\tau^2$ given that the REML estimator is approximately unbiased and efficient. We utilized the *metafor* package in **R** (*78*) to calculate the 95% confidence intervals. We identified significant deviations by noting non-overlaps between estimates on COVID-19 networks and the predicted trends as indicative of significant differences.

Sample non-representativeness is one of the concerns using non-probability samples. To account for potential bias due to the nature of our online sampling, we conducted a raking procedure using the *autumn* package in **R**. This package was developed and used by Democracy Fund + UCLA Nationscape, one of the largest public opinion surveys in the US (*79*). We used monthly CPS data downloaded from the IPUMS website (https://cps.ipums.org/cps/) to construct the target population (*80*). Specifically, we created monthly post-stratified weights to match the marginal distribution against the current population survey (CPS) from April 2020 to May 2022 for the following individual characteristics: sex, age group, race, education, working status, household size, state of residence, metro, survey day from Monday to Sunday, and presidential election voting status (the 2016 presidential election before 2020 November, and the 2020 presidential election on and after 2020 November) in each month. We used post-stratified weights to estimate the overall patterns and trends as well as regression models.

We conducted the extensive set of checks on the performance of the post-stratified weights in multiple ways. Fig. S14 in SI shows that our raking procedure reduces the biases (i.e., the difference in the proportions of the variable between our survey sample and CPS) in most variables used in post-stratification to be close to zero. The mean bias before and after raking across 10 variables and 98 categories is 0.021 and 0.002, respectively. When using the final weight, we showed that the demographics of our weighted sample were similar to those of the general US population, including marital status, which was not part of the raking procedure (Table S6). Comparing COVID-19 vaccine uptake (i.e., the first dose) between our sample estimate and CDC's official rate over time, fig. S2 shows that our sample can generate the 95% confidence intervals of weekly vaccination rates that cover the CDC benchmarks except for the first two weeks in January 2021, where weekly weights were estimated using the same raking procedure that matches sex, age group, race (3 categories; white, Black, Other), education, and household size. Moreover, fig. S3 shows that our data with these weights can make a reasonable prediction on race-specific COVID-19 infection rates. Finally, we calculate various descriptive statistics to characterize core networks with and without post-stratified weights using the whole sample in Table S7. Estimates without weights show slightly larger networks, higher proportions of parent and friendship relationships, a lower proportion of in-person channels, and higher levels of political homophily, but all differences noted here are not substantially large.

**Acknowledgments:** We benefited from comments from Peter Bearman, Delia Baldassarri, Philipp Brandt, Paul DiMaggio, Paula England, Ryan Hagen, Keunbok Lee, Seungwon Lee, Chaeyoon Lim, Kinga Makovi, Mark Pachucki, Barum Park, Alix Rule, David Stark, Daniel Tadmon, Josh Whitford, and Yoosik Youm. An earlier version of paper was presented at the 2022 Population Association of America Annual Meeting, 2022 Politics and Computational Social Science Conference, XLII International Social Networks Conference, the Workshop in Sociology at Indiana University, the CODES seminar on the Center on Organizational Innovation at Columbia University, the Research Group in Network Science Workshop at New York University Abu Dhabi, the Polarization and Radicalization Working Group at New York University, CNS-NRT Speaker Series Colloquium in the Luddy School of Informatics at Indiana University, and the 2023 American Sociological Association Annual Meeting.

**Funding:** This work was supported by the National Science Foundation (#2116936) and the American Assembly at Columbia University.

**Author contributions:**

Conceptualization: BL, KL, BH

Funding acquisition: BL, KL

Project administration: BL, KL

Data Curation: BL, KL

Investigation: BL, KL, BH

Formal Analysis: BL

Visualization: BL

Writing – original draft: BL, KL, BH

Writing - review & editing: BL, KL, BH

**Competing interests:** Authors declare that they have no competing interests.

**Data and materials availability:** All data, codes, and materials to replicate all figures and tables are available at https://doi.org/10.5281/zenodo.10139881.

## Figures

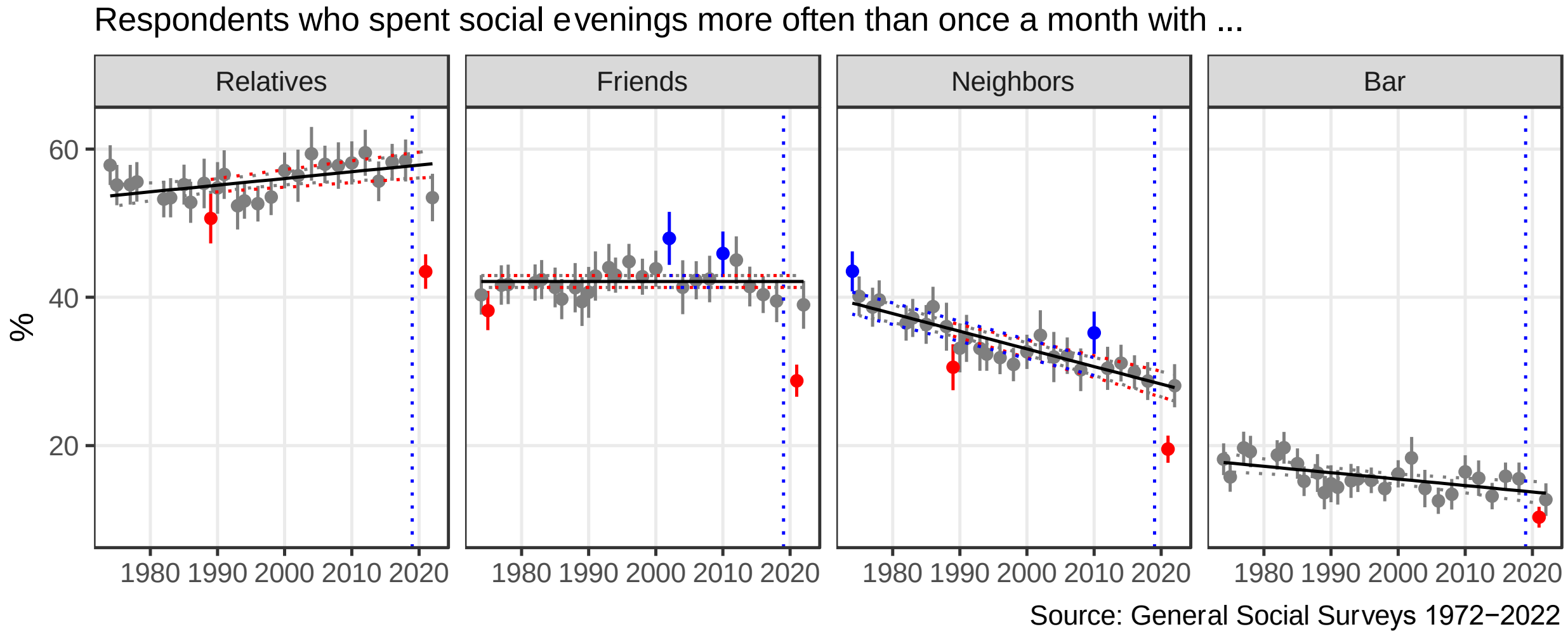

**Fig. 1. Trends in informal social gatherings from 1972 to 2022 in the United States.** Each dot represents the percentage of people who spent a social evening at least once a month with relatives, friends who live outside the neighborhood, someone who lives in your neighborhood, or go to a bar or tavern with 95% confidence intervals. Survey weights are adjusted across the whole analysis. The 95% confidence intervals generated by meta-analysis are used as a benchmark to identify whether estimates on social gathering patterns during the COVID-19 pandemic significantly deviate from the general tendency.

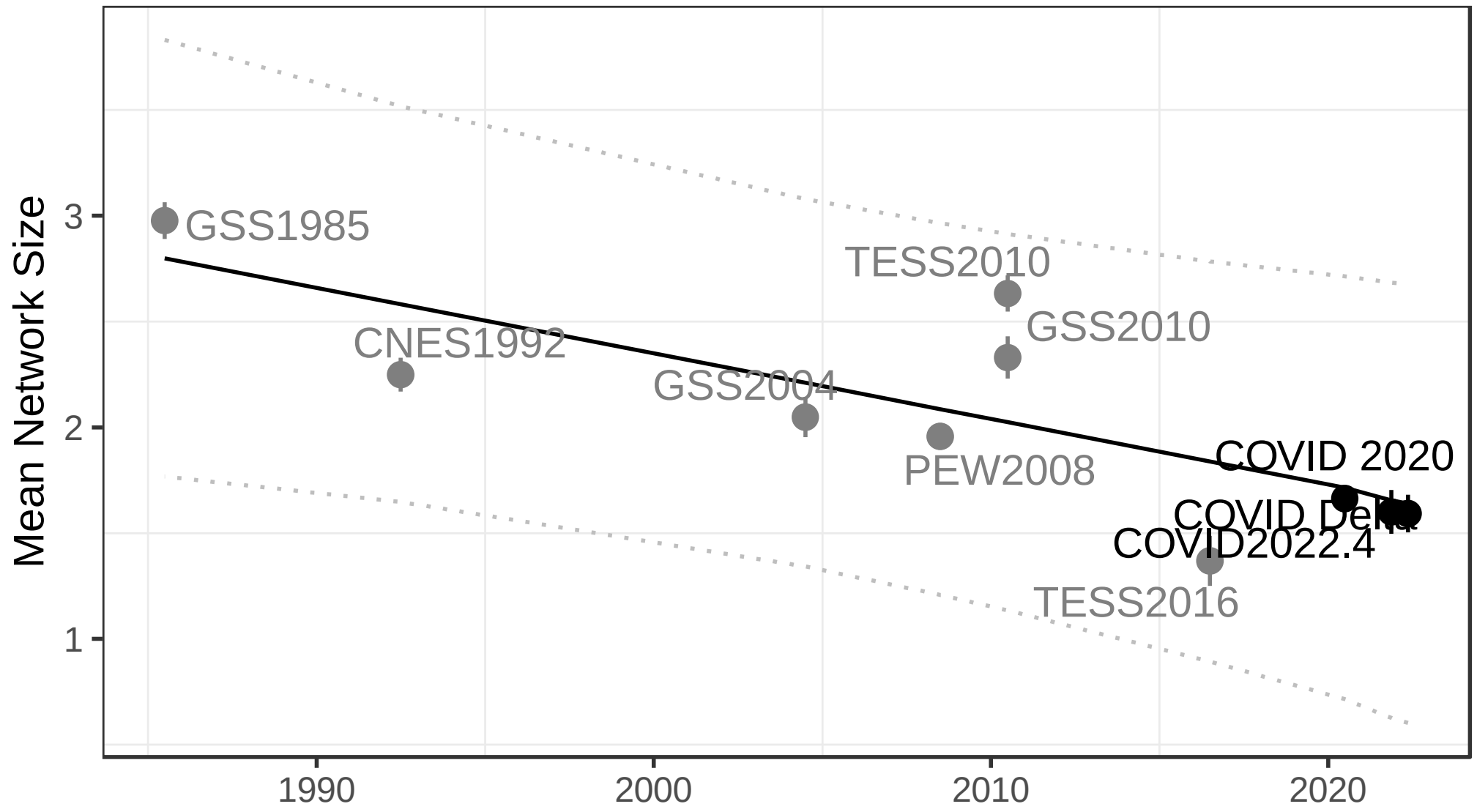

**Fig. 2. The average size of core discussion networks from 1985 to 2022.** Network sizes are capped at five for effective comparison across different surveys (i.e., the maximum network size in the 1992 CNES data was five). Weighted means for network size with 95% confidence intervals are presented. The 95% confidence intervals for average network size in 2020 are very narrow due to the large sample size. The grey box shows the benchmark network size and 95% confidence intervals from the meta-analysis. See Table S1 for the distribution of network sizes as well as their weighted means at various caps in the COVID study and other studies.

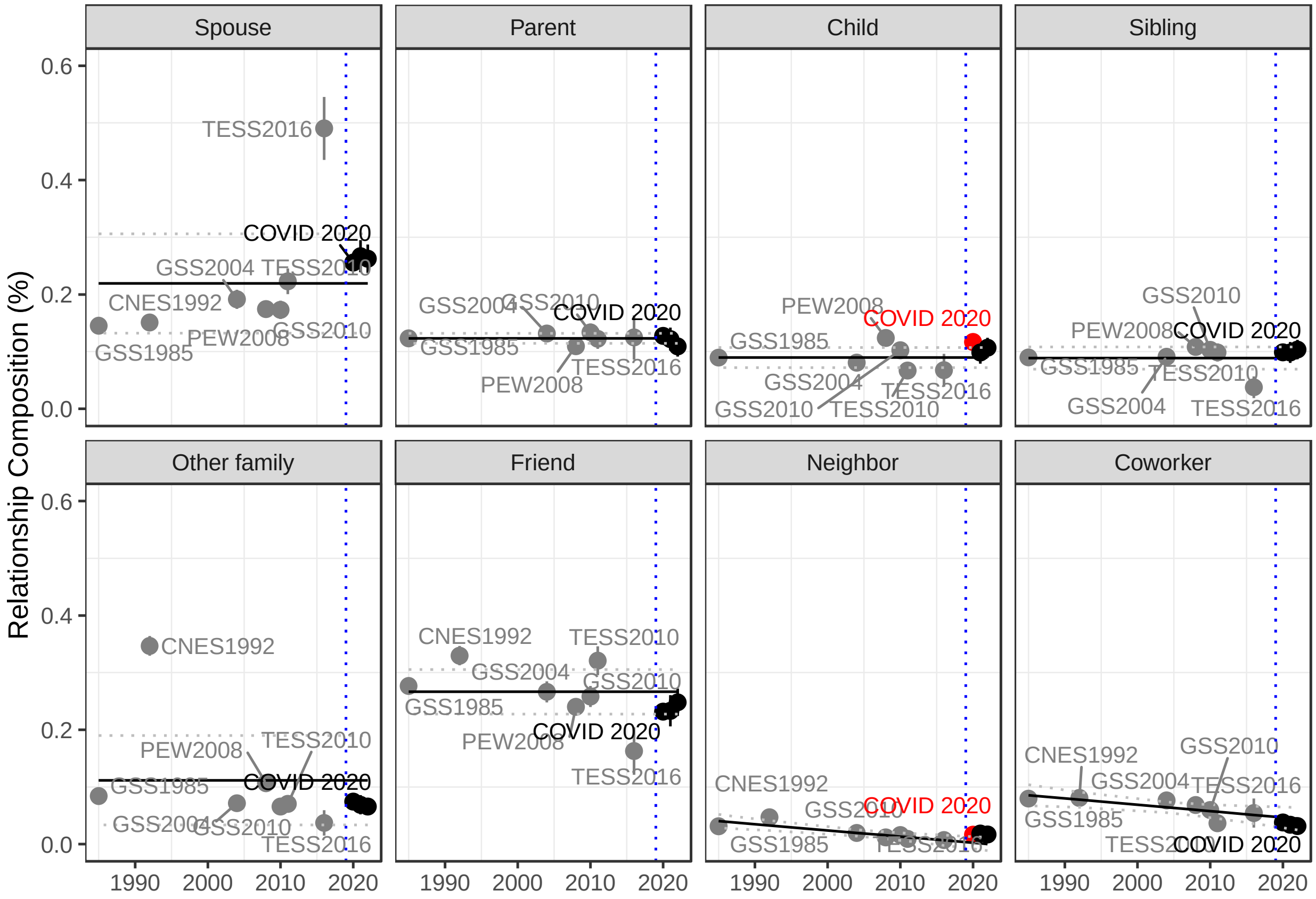

**Fig. 3. The relationship compositions in core discussion networks from 1985 to 2022**. Weighted means for relationship composition with 95% confidence intervals are presented. The "Other" category is omitted here (see Table S2). To account for the fact that multiple responses are allowed for the GSS and CNES studies, we run 1000 random selections of relationship categories and take the average across 1000 runs in 1985, 1992, 2004, 2008 and 2010. Specifically, we use kin-based random selection: first, randomly select one relationship category among kin, and then select one relationship category among other categories, based on the assumption that people would prioritize kin ties over non-kin ties. The grey box shows the mean network size and 95% confidence intervals from a meta-analysis of the network size estimates from 1985 to 2016. Note that the 2016 TESS data only asked about an alter with whom respondents had the last conversation, and second, the 1992 CNES data did only ask whether alters are their spouse or other family without detailed categories.

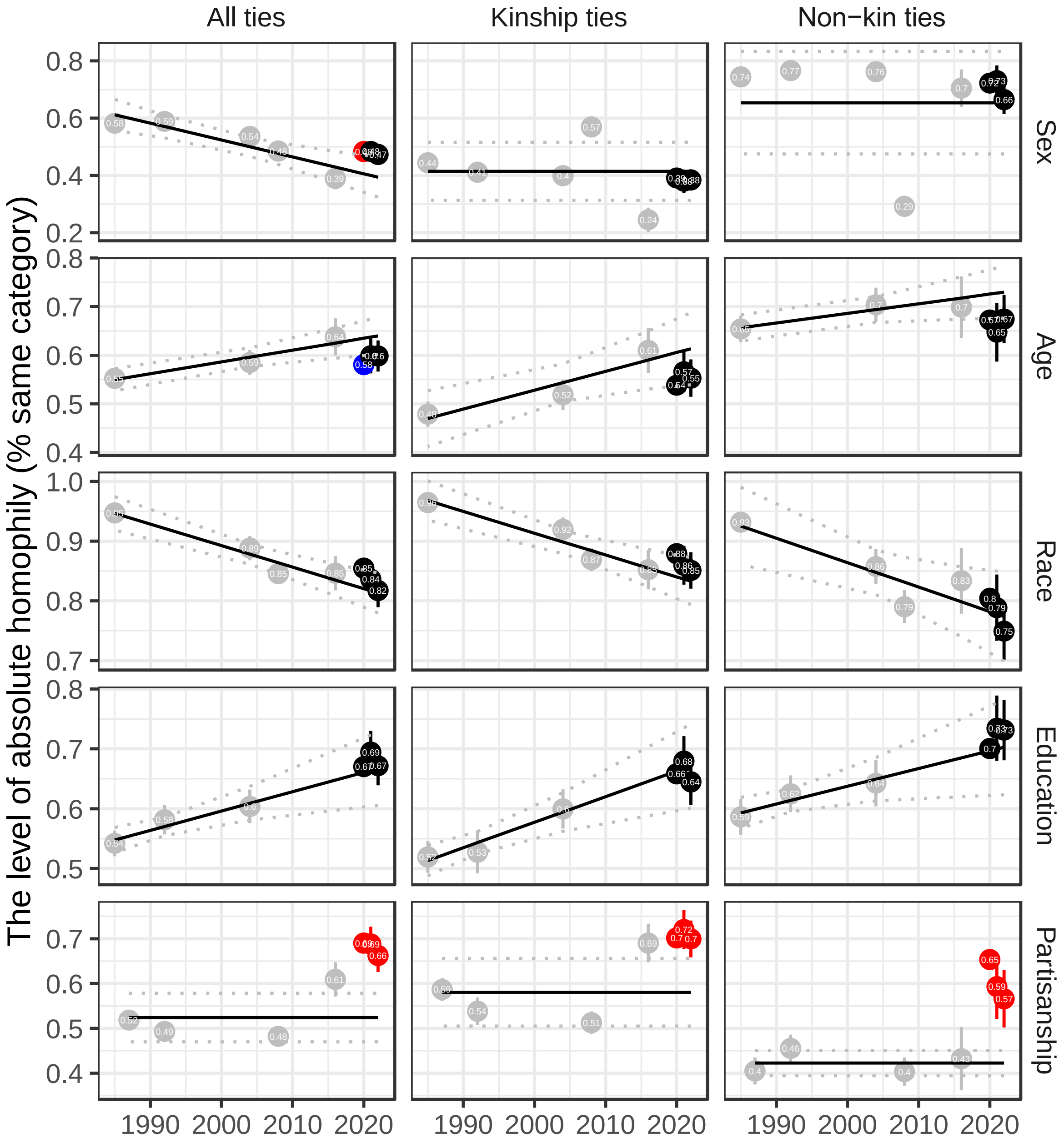

**Fig. 4. The level of absolute homophily across kinship types from 1985 to 2022.** Absolute homophily is measured by the proportion of alters with the same category in the ego's core network. Weighted means for absolute homophily with 95% confidence intervals are presented. The 95% confidence intervals for absolute homophily in 2020 are very narrow due to the large sample size. The grey box shows the mean absolute homophily and 95% confidence intervals from meta-analysis on all available absolute homophily estimates before 2020. Red or blue dots represent when the 95% confidence intervals of absolute homophily from 2020 to 2022 are larger or smaller respectively than those from the 95% confidence intervals on predicted absolute homophily estimates from meta-analysis.

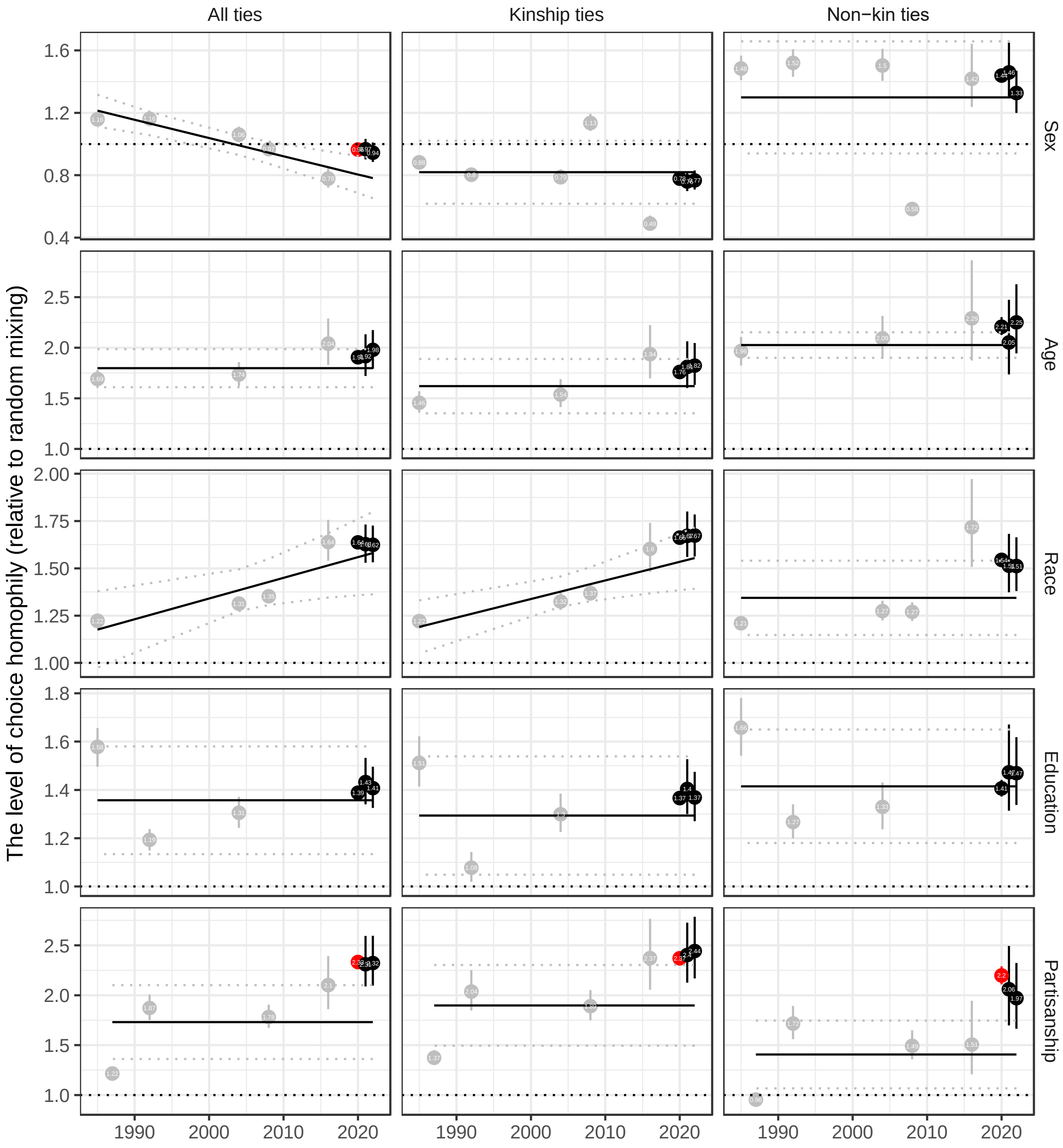

**Fig. 5. The level of choice homophily across kinship types from 1985 to 2022.** Choice homophily is measured by $\frac{\sum p_i}{n}$ / $\left(\frac{\sum_{s=1}^{1000} \frac{\sum p_i^*}{n}}{1000}\right)$, where $p_i$ indicates the proportion of alters with the same category in ego *i*'s core network and $p_i^*$ indicates the proportion from simulated ego-networks that fix the degree distribution and population composition. Given that choice homophily is measured by the ratio of absolute homophily over chance homophily, we can identify homophilious relationships if it is significantly larger than one (i.e., black dotted line), and heterophilious relationships if it is below one. Weighted means for choice homophily with 95% confidence intervals are presented. The 95% confidence intervals for choice homophily in 2020 are very narrow due to the large sample size. The grey box shows the mean choice

homophily and 95% confidence intervals from meta-analysis on all available choice homophily estimates before 2020. Red dots represent when the 95% confidence intervals of choice homophily from 2020 to 2022 are larger than that from the 95% confidence intervals from meta-analysis.

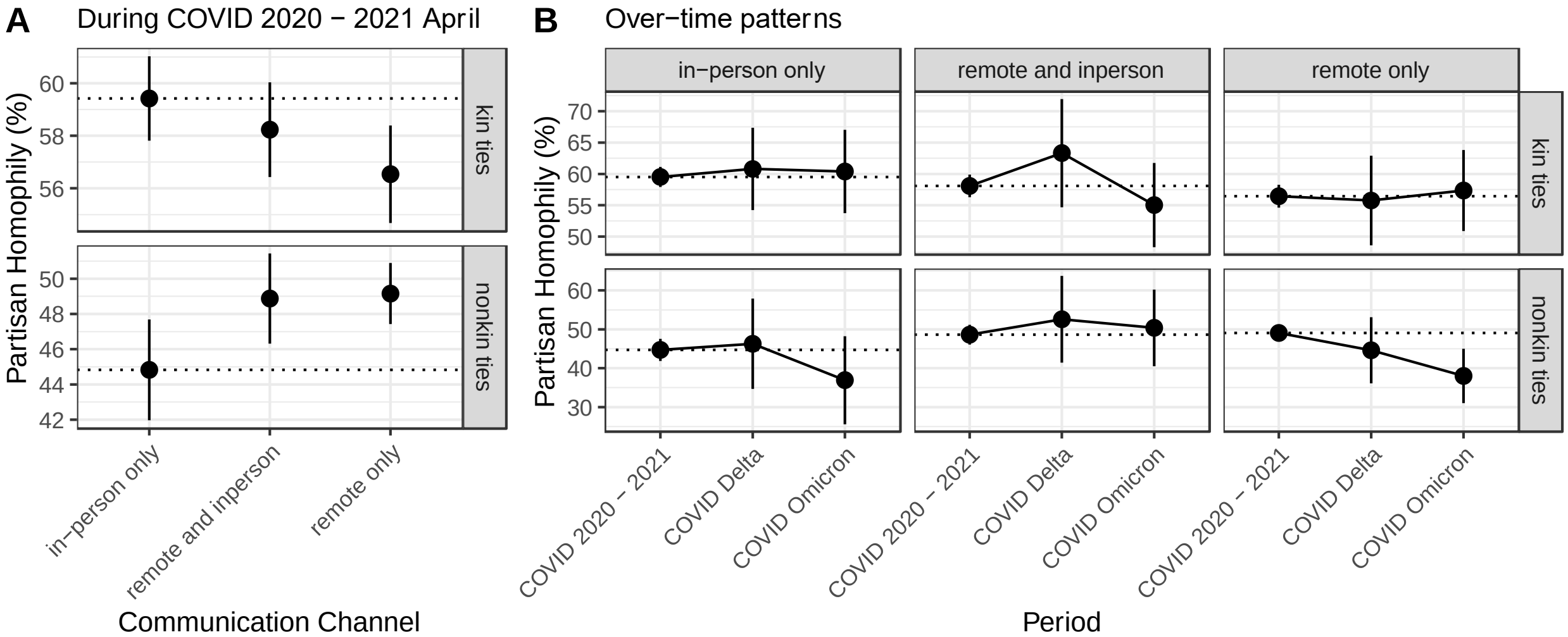

**Fig. 6. The role of remote communication channels in the level of political homophily**. We estimate logistic regression models to predict political homophily (=1 if an ego and an alter share the same partisanship, 0 otherwise) after accounting for individual confounders and the location of alters. Each dot indicates predicted margins with 95% confidence intervals of political homophily across communication channels and relationship types. Filled dots show when the average marginal effects (ref = in-person only in the left panel and estimates from COVID 2020-2022 in the right panel) are statistically significant at p = 0.05; otherwise, they are empty.

Supplementary Materials for

# Transformation of social relationships in COVID-19 America: Remote communication may amplify political echo chambers

Byungkyu Lee* et al.

*Corresponding author. Email: bklee@nyu.edu

**This PDF file includes:**

Supplementary Text
Figs. S1 to S14
Tables S1 to S7

## Supplementary Text

Appendix A. Pretest on network name generators

In online network surveys, the number of boxes presented in the network name generators can affect the number of names provided due to individuals' motivation to "fill in" all the boxes. To avoid such satisfying and heaping effects, existing 2010 and 2016 TESS studies used a one-box design, though this design may take longer as it involves repeating the same questions, "Anyone else?" We conduct a survey experiment as part of the pretest to examine how the number of boxes affects network size, isolation, relationship type, and survey duration, before implementing our main survey.

We randomly assigned respondents to one of six experimental conditions. In the one-box design, we first asked whether they discuss "personally important matters" with others, and then we asked whether they discuss health problems and political matters with others separately. While this design allowed us to differentiate between alters invoked by different name generators, it could be time-consuming. In the multi-box design, we asked whether they discuss "important matters" with others, nudging that important matters may include personal, political, and health-related problems. While this multi-box design could be time efficient, it was difficult to distinguish how they discuss three different matters with the same or different alters. If respondents said they had someone to discuss these matters with, we asked them to provide the first names or initials of their alters. In the one-box design, we asked the follow-up question, "Anyone else?" until we collected five names for the important matters name generators, and three names for the political and health matters name generators, respectively. In the multi-box design, we present two, three, five, seven, or ten boxes, depending on the different experimental conditions, with a one-time follow-up question, "Anyone else"?

Fig. S1, Panel A shows a positive association between the number of boxes presented and network size, but the network size from the one box design is similar to those from the other designs. Panel B shows that the number of boxes does not significantly affect the level of isolation. Panel C shows that the duration of surveys in the one-box design is not significantly different from those from other designs. Panel D confirms that the distribution of relationship types collected by the one-box design is similar to those from other designs, except for the higher proportion of spouse in the seven-box design, the higher proportion of parent in the two-box design, and the higher proportion of coworker in the ten-box design.

Common prompt across all conditions:
Now we are going to ask you some questions about your relationships with other people. We will begin by identifying some of the people you interact with on a regular basis. You may refer to these people in any way you want; for example, you may use just their first names or initials. We are not interested in the identities of these people. We just need to have some way to refer to them so that when we ask you some follow-up questions, we both know whom we are talking about.

In one box design, respondents are first presented with the personally important matters name generator:

Q1. From time to time, most people discuss personally important matters with other people. Looking back over the last month – who are the people with whom you discussed matters personally important to you? Do you have anyone?

Q2. (If Q1 is yes) Please type the first name or initial of one person below. You'll have the chance to tell us about additional people you talk to in the next few questions.
______________________________________________________________

If respondents have provided some names, then we ask

Q3. Anyone else?
Yes. Add additional name in the box : _________
No. I have no more names to add.

Repeat Q3 until it collects the five names.

The presentation of health matters (Q4) and political matters (Q5) name generators is randomly ordered.

Q4. From time to time, most people discuss health problems with others. Looking back over the last month, who are the people with whom you discuss your physical, mental, and emotional health matters? Do you have anyone?

Q5. From time to time, most people discuss government, elections, and politics with others. Looking back over the last month, who are the people with whom you discuss political matters? Do you have anyone?

If respondents do not mention one person before,

Q6-1. Please type the first name or initial of one person below. You'll have the chance to tell us about additional people you talk to in the next few questions.
______________________________________________________________

If respondents mention at least one person before,

Q6-2. Please type the first name or initial of one person below. He or she may be the same or different from the people on the list you provided earlier. If he or she is the same person, please type the same first name or initial.
______________________________________________________________

If respondents have provided some names, then we ask:

Q8. Anyone else?
Yes. Add additional name in the box : _________
No. I have no more names to add.

Repeat Q8 until it collects the three names for each name generator.

In the multi-box design, respondents are first presented with the general important matters name generator.

Q1. From time to time, most people discuss important matters with other people. They may include personal matters, politics, and/or health-related problems. Looking back over the last month – who are the people with whom you discussed matters important to you? Do you have anyone?

Q2. Who are those people with whom you discussed matters important to you? Please type their first names or initials below, one person in a box.

| N box | 2 | 3 | 5 | 7 | 10 |
|---|---|---|---|---|---|
| Response Options | | | | | ___________ |

If respondents have filled all boxes, then we ask
Q3. Anyone else?
Yes. Add additional name in the box : _________
No. I have no more names to add.

Panel A. Network size

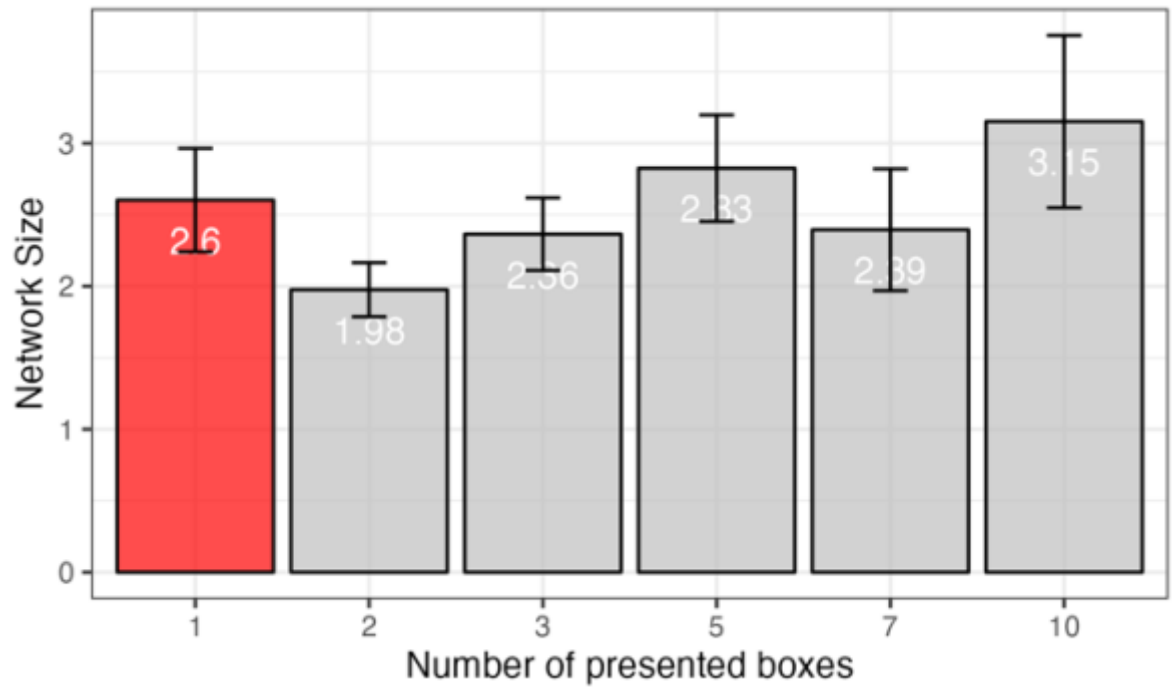

Panel B. Network isolation

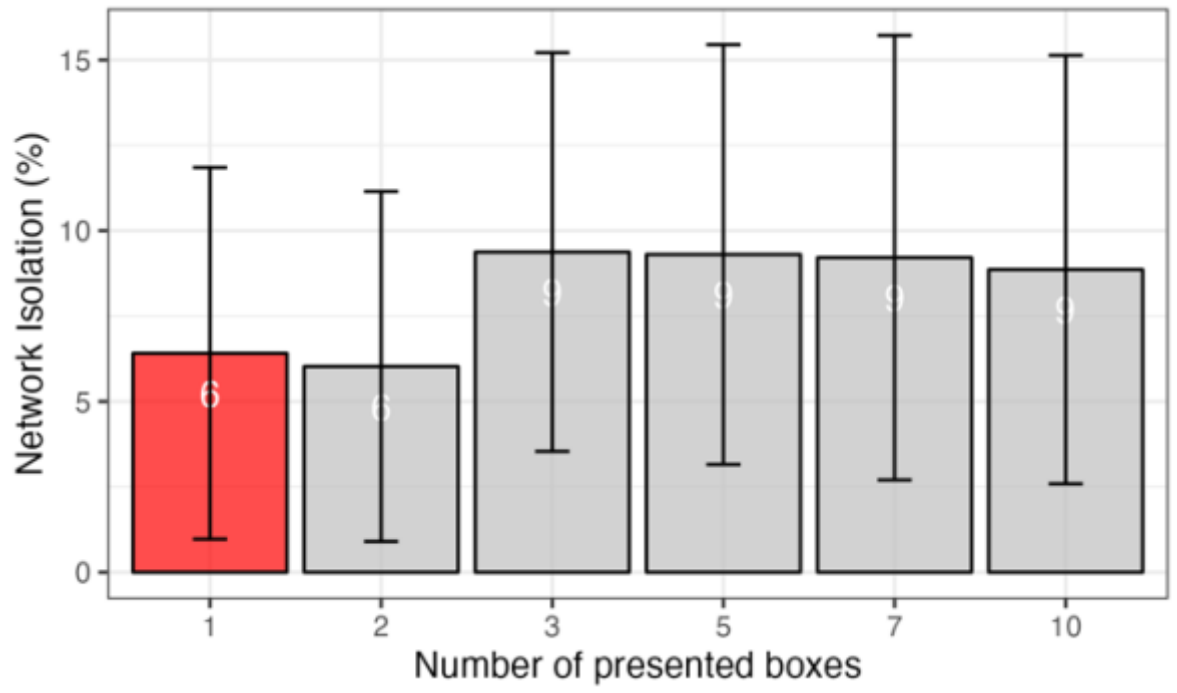

Panel C. Survey length

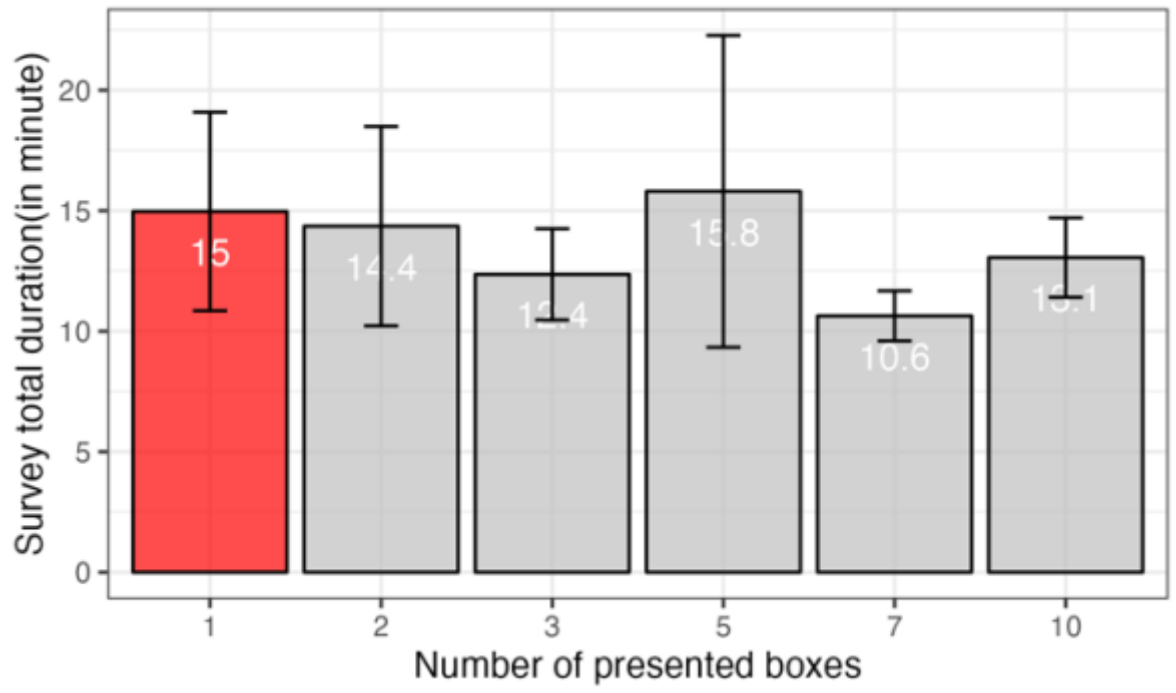

Panel D. Relationship Type

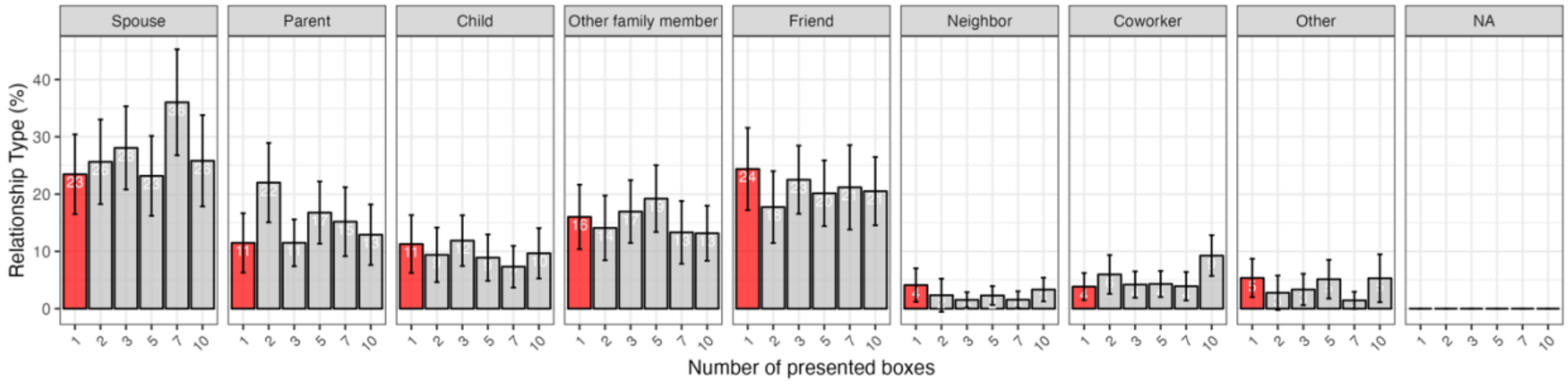

**Fig. S1. The comparison of network size, network isolation, survey length, and relationship type by the number of presented boxes.**

Appendix B. Introduction to network name generators and name interpreters in the COVID-19 network study and other surveys.

The ego-centric network survey consists of two parts: name generator and name interpreter. The content of network name generators is crucial since it determines which ties people would report as part of their core discussion networks. The earliest nationally representative ego-centric network survey was the General Social Survey in 1985 which uses “important matters” name generators. Since then, numerous surveys have followed a similar format with slightly different wordings (see the bottom of Appendix F for the exact wordings of all network name generators in all eight studies). Based on the pretest results on the number of boxes in network name generators (see Appendix B), we decided to use the multiple generators with a one-box design instead of a single-generator multi-box design. The former, mimicking traditional offline network surveys, enables us to examine tie multiplexity as well.

Here, we first introduce our network name generators, followed by network name interpreters. In our 2020 COVID survey, we employed three name generators that asked with whom people discuss (a) important matters, (b) political matters, and (c) health matters. We asked respondents to type a name (e.g. initials, first names, nicknames) in a box for each network name generator. After collecting up to five names that respondents discuss personally important matters with, we asked them to provide a name in a box again for political matters and health matters name generators, in random order. We specifically ask them to type the same name if they refer to the same person. If they indicated that they had no one to discuss, we then asked why for each network name generator. After collecting all names, we created a unique name list per respondent using Qualtrics’ JavaScript and asked about the nature of relationships and characteristics of the alters using the name interpreters. Specifically, the 2020 COVID network survey asked:

Q1. Now we are going to ask you some questions about your relationships with other people. We will begin by identifying some of the people you interact with on a regular basis. You may refer to these people in any way you want; for example, you may use just their first names, nicknames, initials, or relationships with you. We are not interested in the identities of these people. We just need to have some ways to refer to them so that when we ask you some follow-up questions, we both know whom we are talking about.

Respondents are first presented with the personally important matters name generator.

Q2. From time to time, most people discuss personally important matters with other people. Looking back over the last month – who are the people with whom you discussed matters personally important to you? Do you have anyone?

(If Q2 is Yes), Q3. With whom did you discuss important matters? Please type the first name, nickname, or initial of one person below. You’ll have the chance to tell us about additional people you talk to in the next few questions.

________________________________________________________________

If respondents have provided some names, then we ask,

Q4. Anyone else?
Yes. Add additional name in the box : _________
No. I have no more names to add.

Repeat Q4 until it collects the five names.

The presentation of the following health matters (Q5) and political matters (Q6) name generators is randomly ordered.

Q5 From time to time, most people discuss health-related matters with others. Looking back over the last month, who are the people with whom you discussed your physical, mental, and emotional health? Do you have anyone?

Q6. From time to time, most people discuss government, elections, and politics with others. Looking back over the last month, who are the people with whom you discussed political matters? Do you have anyone?

If respondents do not mention one person before,

Q7-1. With whom did you discuss health-related matters / political matters? Please type the first name, nickname, or initial of one person below. You'll have the chance to tell us about additional people you talk to in the next few questions.
________________________________________________________________

If respondents mention at least one person before,

Q7-2. With whom did you discuss health-related matters / political matters? Please type the first name, nickname, or initial of one person below. He or she may be the same or different from the people on the list you provided earlier. If he or she is the same person, please type the exact same name you used before.
________________________________________________________________

If respondents have provided some names, then we ask,

Q8. Anyone else?

Yes. Add additional name in the box : _________
No. I have no more names to add.

Repeat Q8 until it collects the three names.

Once we collect all names, then we ask the following set of network name interpreters (Q10-Q17) after expressing our thanks.

Q9. Thanks for providing the names of people with whom you discuss matters that are either personally important, health-related, or political. We will ask you about the nature of your relationship with them, and their demographic characteristics. Keep in mind that your answers to these questions will be kept confidential and anonymous.

Q10. Thinking back to the most recent discussion you had with them: what did you talk about? Please check all that apply.

| Family, Kids and Education | Work and Career | Personal Finance, and Housing | Health | Relationship Issues | Politics | Other topic | Don't remember |
|---|---|---|---|---|---|---|---|

Q11. Below is a list of ways that people can be connected to each other. Which of the following best describes your relationship with them? Please choose the only one that best describes each relationship.

| Parent | Spouse | Sibling | Child | Other family member | Co-worker | Friend | Neighbor | Other |
|---|---|---|---|---|---|---|---|---|

Q12. Over the last month, have you discussed things that are related to Coronavirus (i.e., COVID-19) with each person?

| No | Yes | Don’t remember |
|---|---|---|

Q13. What did you use for your communication the last time you talked to each person? Please check all that apply.

| In-person | Telephone | Video Call (e.g., Skype) | Text Message (e.g., WhatsApp) | Email | SNS (e.g., Facebook, twitter) | Other |
|---|---|---|---|---|---|---|

Q14. What is each person's sex?

| Male | Female |
|---|---|

Q15. What is each person's race? Please choose the only one that best describes each person's race.

| White | Black | Hispanic | Asian | Other | Don't know |
|---|---|---|---|---|---|

Q16. How old is each person?

| Younger than 20 | 20-39 | 40-59 | Older than 60 | Don't know |
|---|---|---|---|---|

Q17. What is the highest level of education each person has completed?

| Less than high school | High school degree or GED | Some college | College degree or higher | Don't know |
|---|---|---|---|---|

Q18. Generally speaking, does each person probably think of himself/herself as a Republican, Democrat, Independent, or Something else?

| Republican | Democrat | Independent | Something else | Don't know |
|---|---|---|---|---|

The 1985, 2004, 2010 General Social Surveys, and the 2010 Time-Sharing Experiment for Social Sciences Study asked:

> "From time to time, most people discuss important matters with other people. Looking back over the last six months—who are the people with whom you discussed matters important to you? Just tell me their first names or initials." IF LESS THAN 5 NAMES MENTIONED, PROBE: "Anyone else?"

The 1992 Cross-National Election Studies asked:

> "From time to time, most people discuss important matters with other people. Looking back over the last six months, I'd like to know the people you talked with about matters that are important to you. Can you think of anyone?" IF LESS THAN 4 NAMES MENTIONED, PROBE: "Is there anyone else you talk with about matters that are important to you?" [AFTER THE SECOND NAME: "Anyone else (you can think of)?"] AFTER THEN, ASKING: "Aside from[n] anyone you have already mentioned, who is the person you talked with most about the events of the recent presidential election campaign?"

The 2008 PEW study asked:

> "From time to time, most people discuss important matters with other people. Looking back over the last six months — who are the people with whom you discussed matters that are important to you? If you could, just tell me their first name or even the initials of their first AND last names." [RECORD UP TO 5 - LOOP. IF LESS THAN 5 MENTIONED, PROBE: "Anyone Else?"] [INTERVIEWER: If R says a relationship instead of a name/initials (e.g. "my wife" or "my brother"), change "my" to "your" – i.e. type as "your wife" or "your brother"] [PROGRAMMER: Also include "No additional mentions" (Code 97), "Don't know" (98) and "Refused" (99) punches]

The 2016 Time-Sharing Experiment for Social Sciences Study asked:

> "From time to time, most people discuss government, elections and politics with other people. Looking back over the last month – who are the people with whom you discussed matters political to you? Just tell us their first names or initials." IF NEITHER OF CHECK BOXES IS CHECKED, DO NOT ALLOW TO PROCEED UNTIL 1 OF THEM IS CHECKED. SHOW THE FOLLOWING PROMPT: "Please check "Add another name" if you would like to add more names or check "I have no more names to add" if you do not want to add any more names."

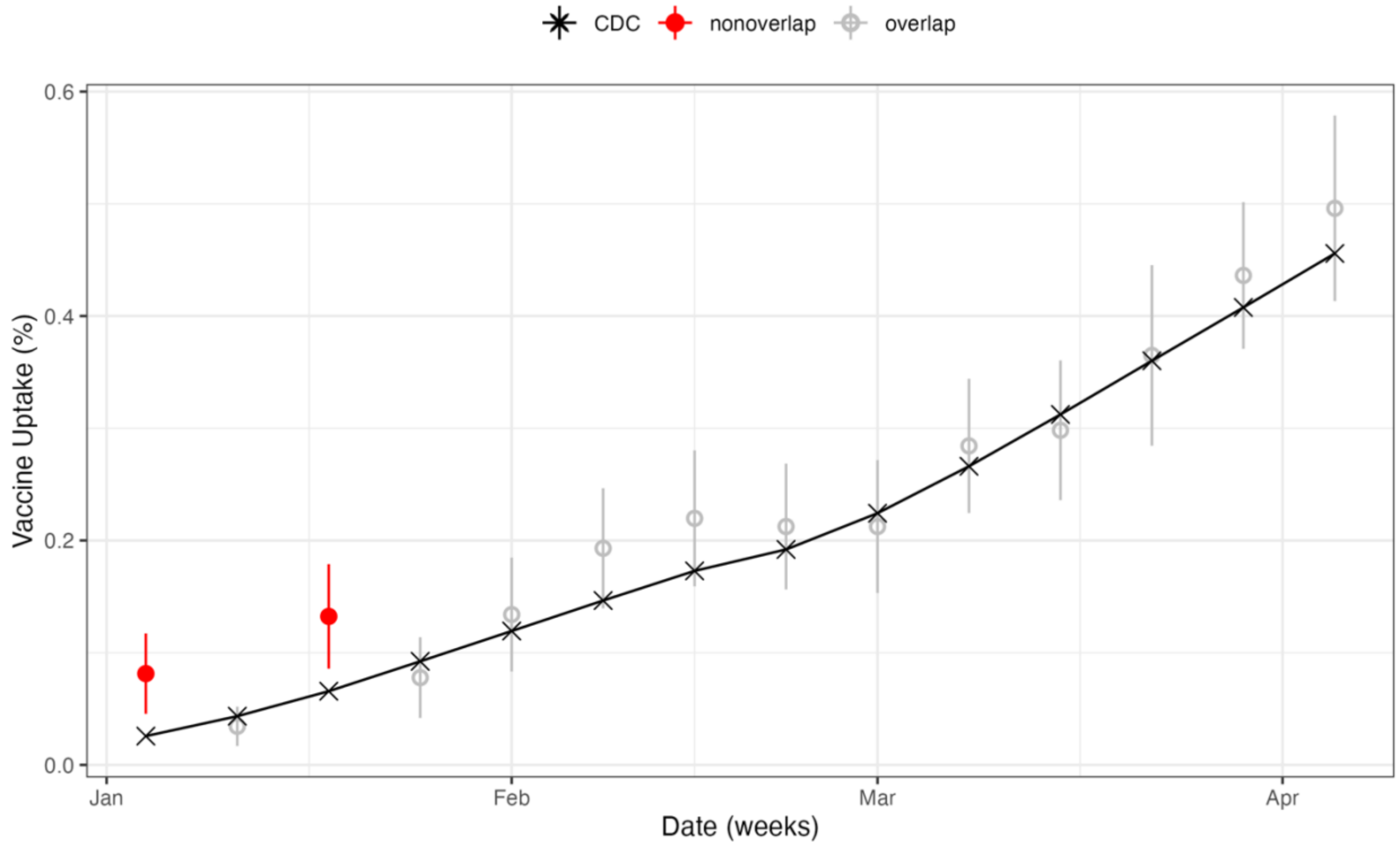

**Fig. S2. Estimated COVID-19 vaccination rates from our survey against the CDC's official vaccination rates.** Each circle dot represents the estimated weekly vaccination rates with 95% confidence intervals from our survey from 2021 January to 2021 March, adjusted by post-stratified weights derived from survey raking. Each X denotes the weekly vaccination rate from CDC. Red dots indicate when our survey-based estimates failed to predict CDC benchmark vaccination rates with 95% confidence intervals, and grey dots indicate when they overlap.

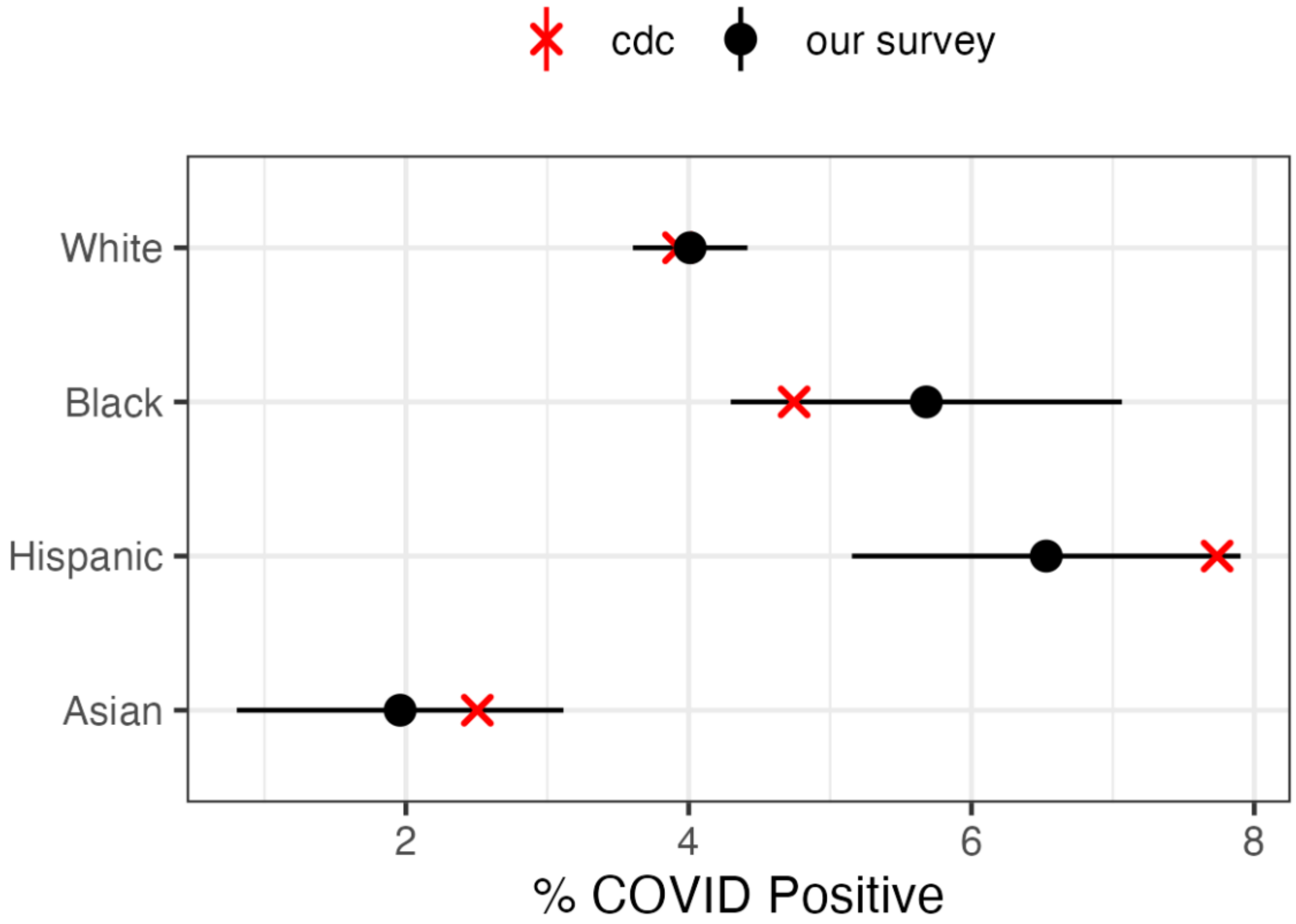

**Fig. S3. Estimated race-specific COVID-19 infection rates from our survey against the CDC's official COVID infection rates.** Each circle dot represents the estimated COVID-19 vaccination rates with 95% confidence intervals from our survey from 2020 April to 2021 March, adjusted by post-stratified weights derived from survey raking. Each X denotes the COVID-19 positivity rates from CDC during the same period.

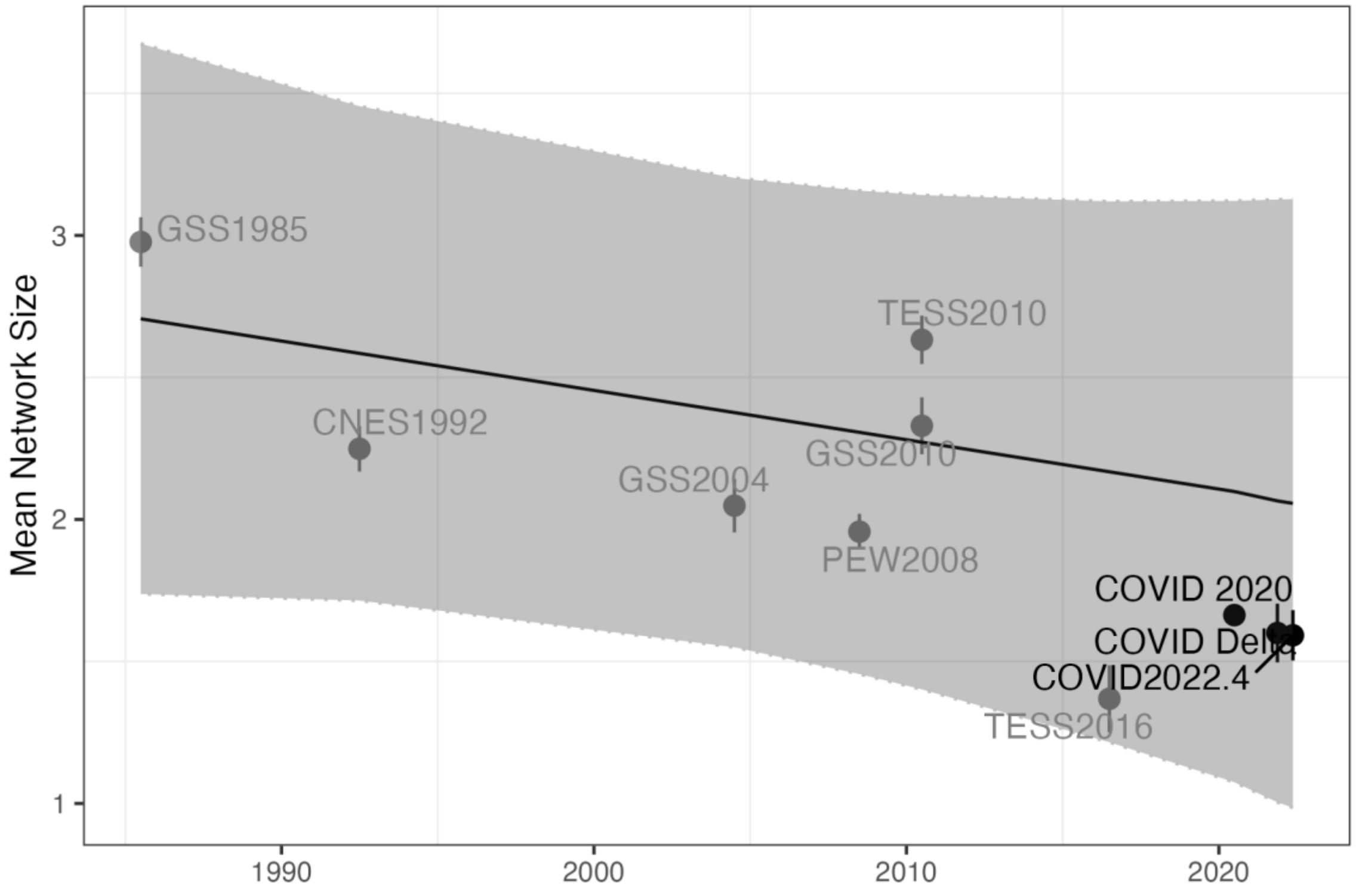

**Fig. S4. The trends in the size of core discussion networks from 1985 to 2022 while excluding the 2004 GSS and 2016 TESS studies in the benchmark.** Network sizes are capped at five for effective comparison across different surveys (i.e., the maximum network size in the 1992 CNES data was five). Weighted means for network size with 95% confidence intervals are presented. The 95% confidence intervals for average network size in 2020 are very narrow due to the large sample size. The grey box shows the benchmark network size and 95% confidence intervals from the meta-analysis that exclude the 2004 GSS and 2016 TESS data to assess the impact of outliers.

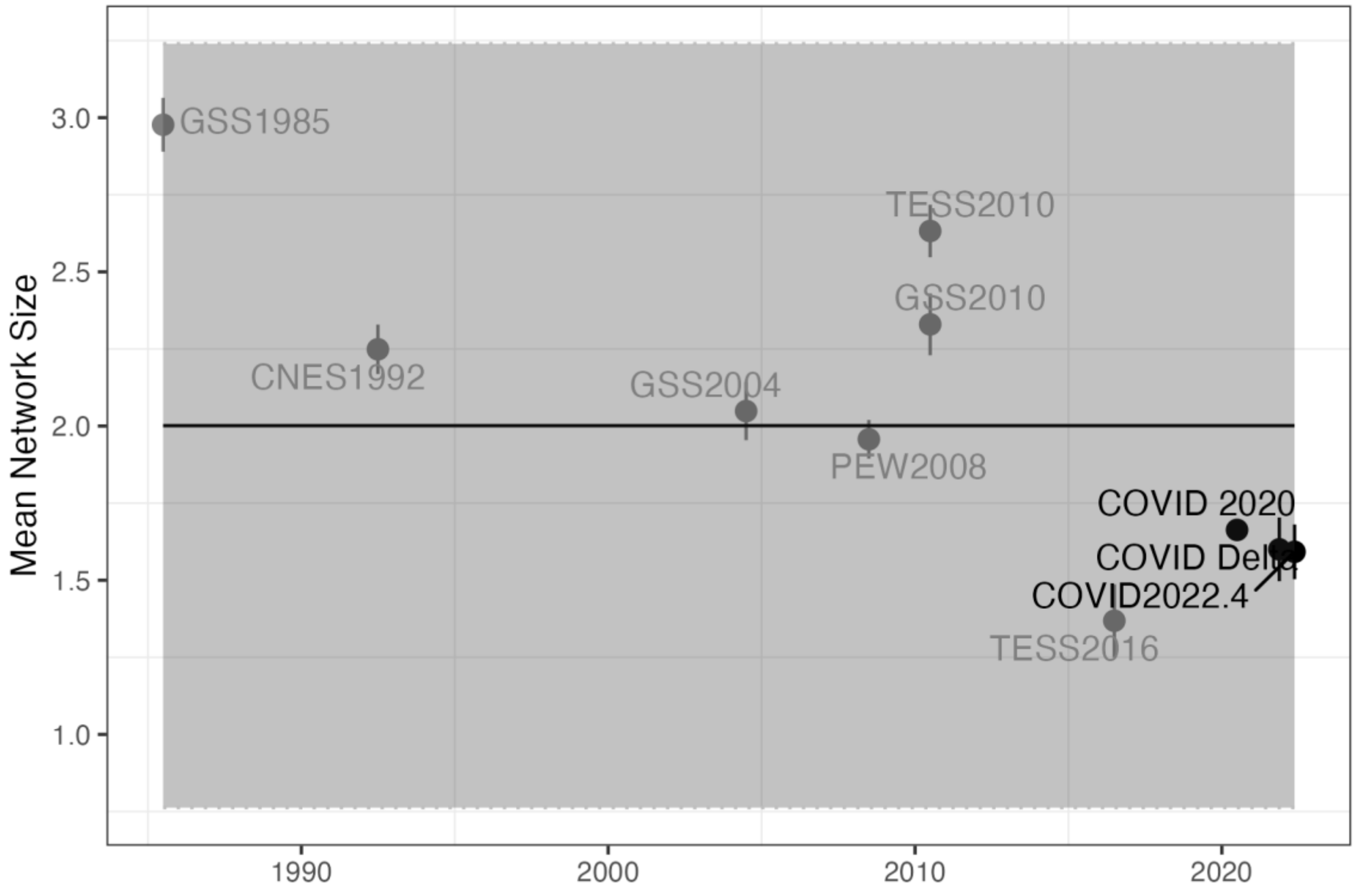

**Fig. S5. The trends in the size of core discussion networks from 1985 to 2022 while only using the 2010 and 2016 TESS studies in the benchmark.** Network sizes are capped at five for effective comparison across different surveys (i.e., the maximum network size in the 1992 CNES data was five). Weighted means for network size with 95% confidence intervals are presented. The 95% confidence intervals for average network size in 2020 are very narrow due to the large sample size. The grey box shows the benchmark network size and 95% confidence intervals from the meta-analysis including only the 2010 and 2016 TESS surveys that use the same web-based design with the COVID network survey. As it is not feasible to include a linear trend term in the meta-analysis with only two observations, we calculate the mean and 95% confidence intervals for the benchmark.

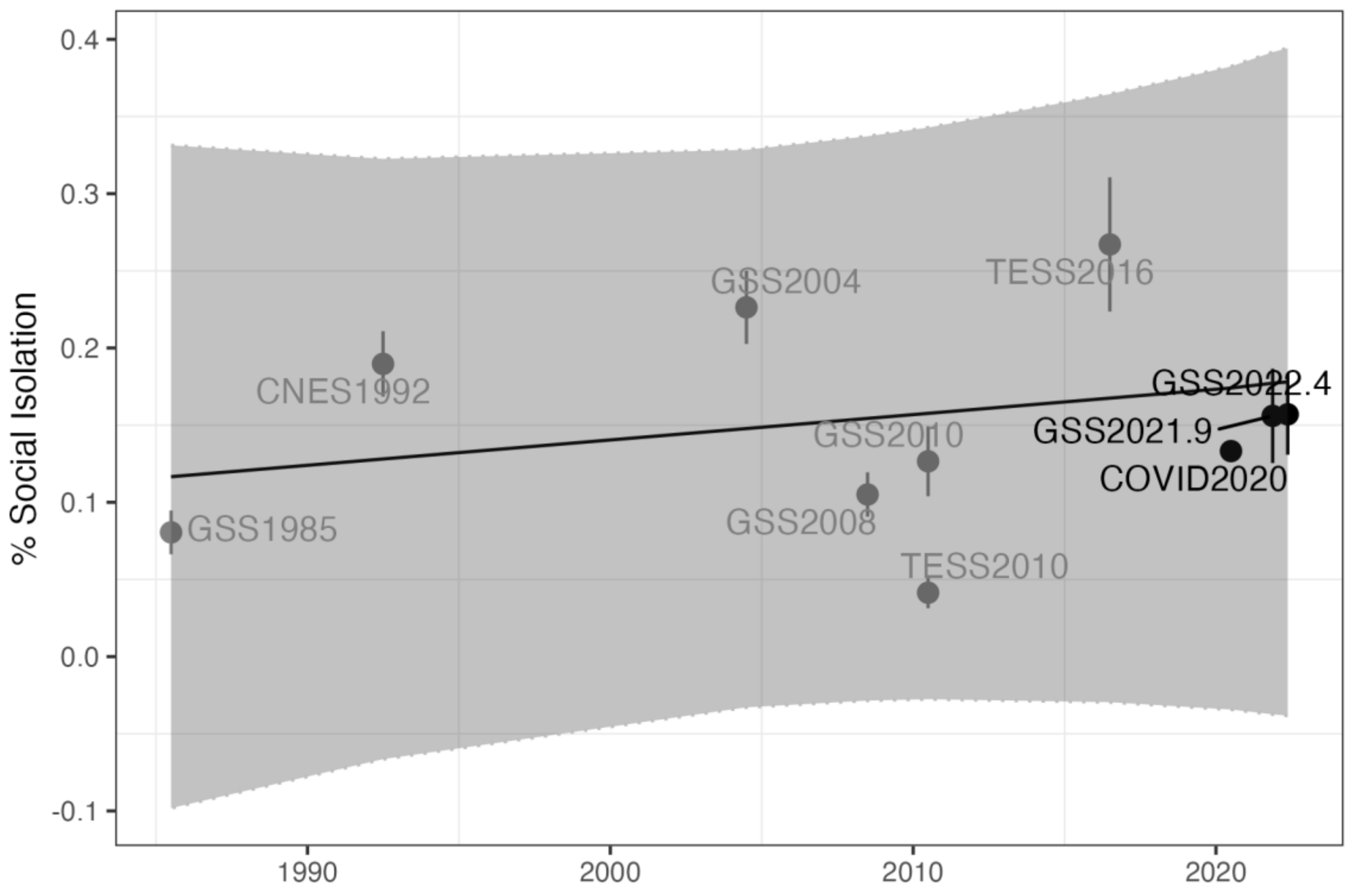

**Fig. S6. Trends in social isolation from 1985 to 2022.** Weighted means for social isolation with 95% confidence intervals are presented. The 95% confidence intervals for social isolation in 2020 are very narrow due to the large sample size. The grey box shows the mean social isolation and 95% confidence intervals from a meta-analysis of the social isolation estimates from 1985 to 2016.

Panel A. The distribution of race category

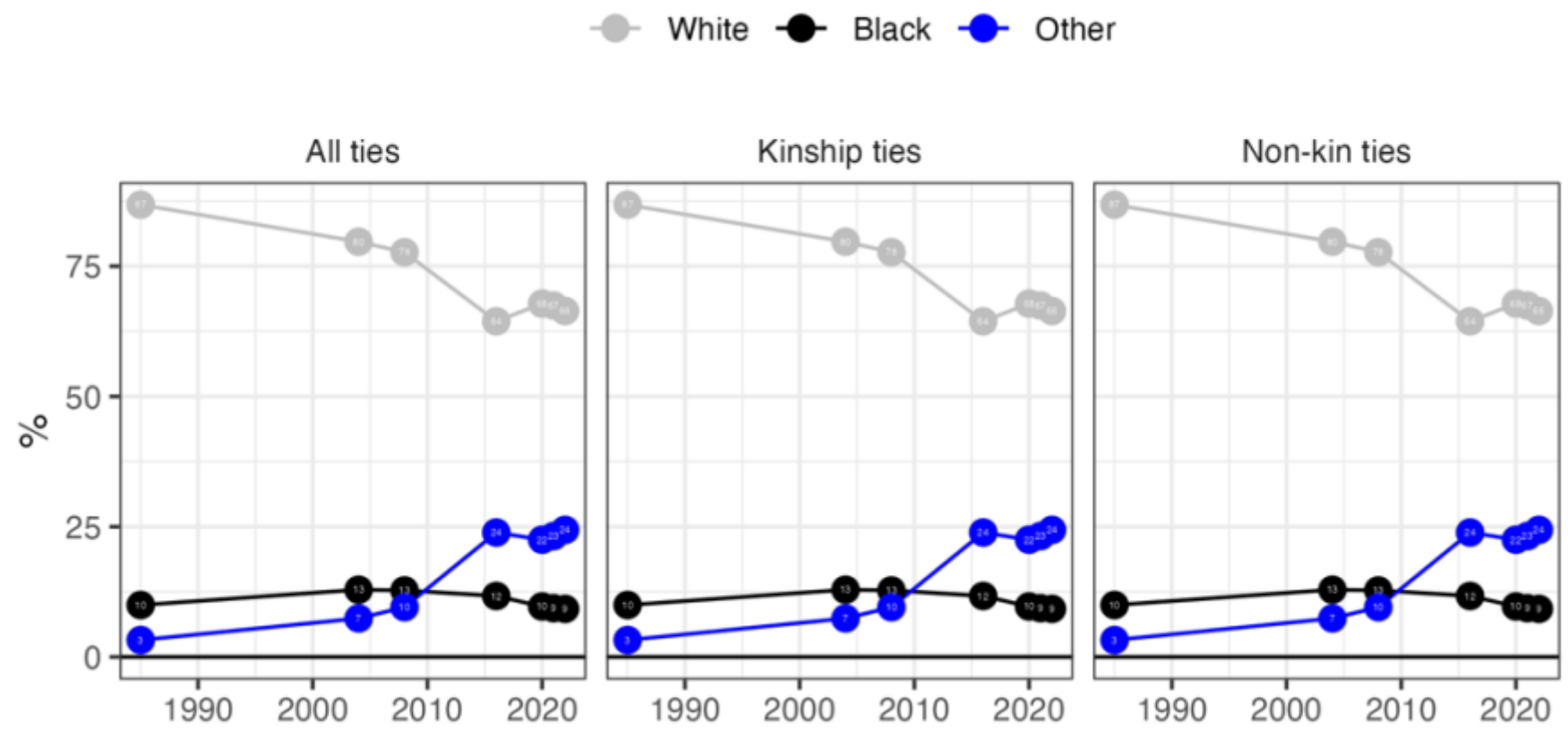

Panel B. The level of absolute homophily: the % same race

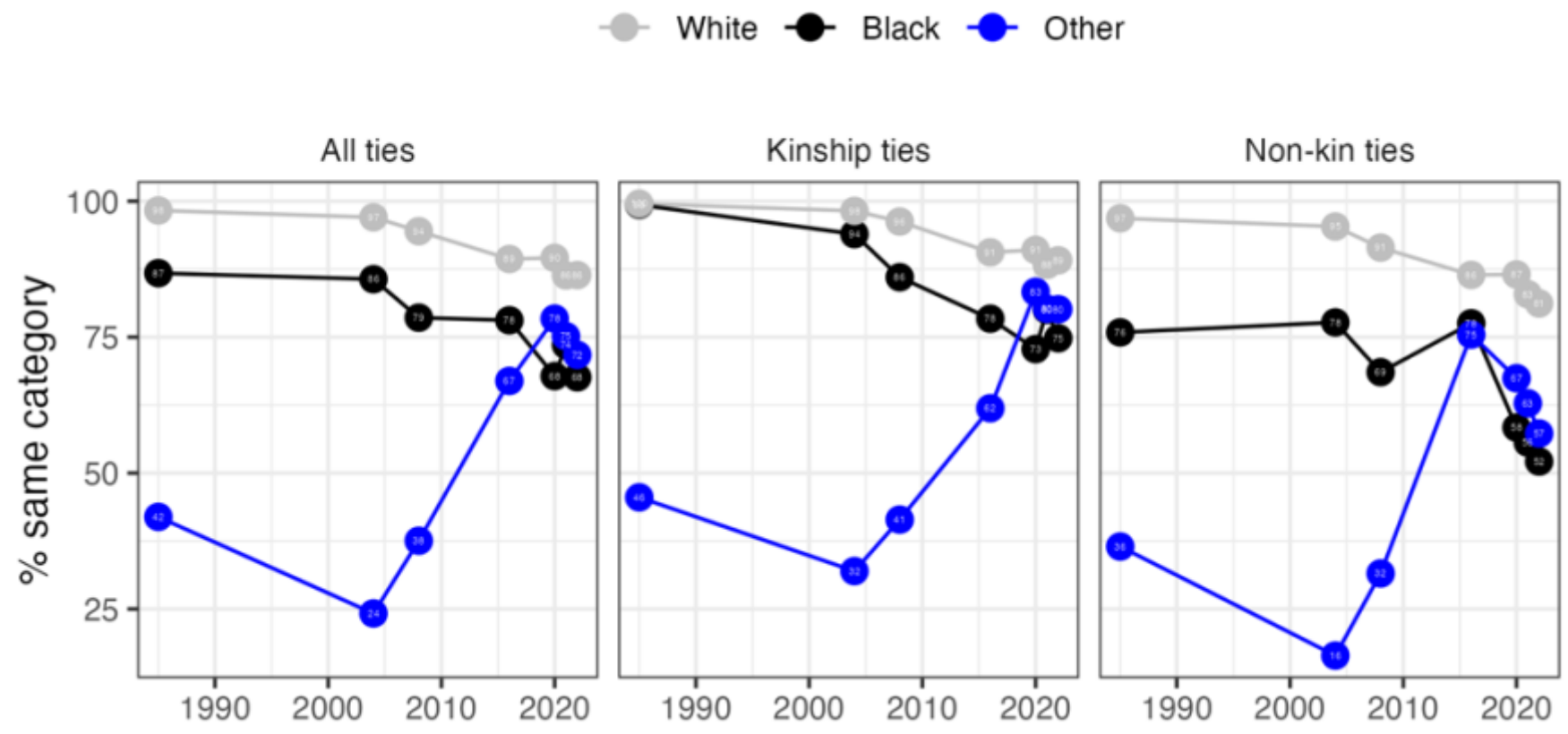

Panel C. The level of choice homophily: Coleman index: from -1 (heterophily) to 1 (homophily)

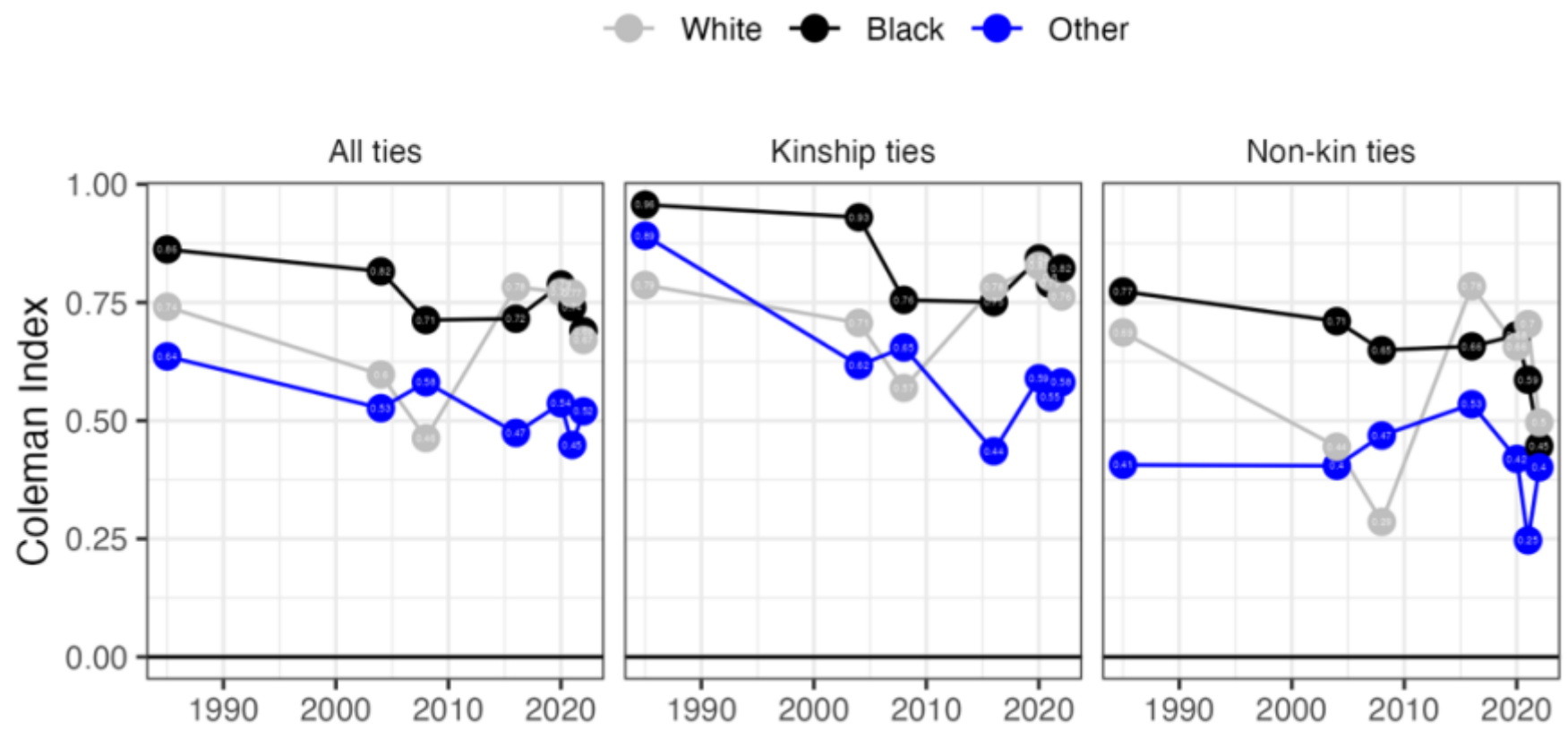

**Fig. S7. The decomposition of racial homophily across different racial groups from 1985 to 2022.** Panel A shows the distribution of each race category over time, which has been used to control for structural opportunities in calculating the Coleman index (Panel C). Panel B shows the level of absolute homophily measured by the proportion of the same race ties. Panel C shows the level of choice homophily measured by the Coleman index.

Panel A. The distribution of partisanship category

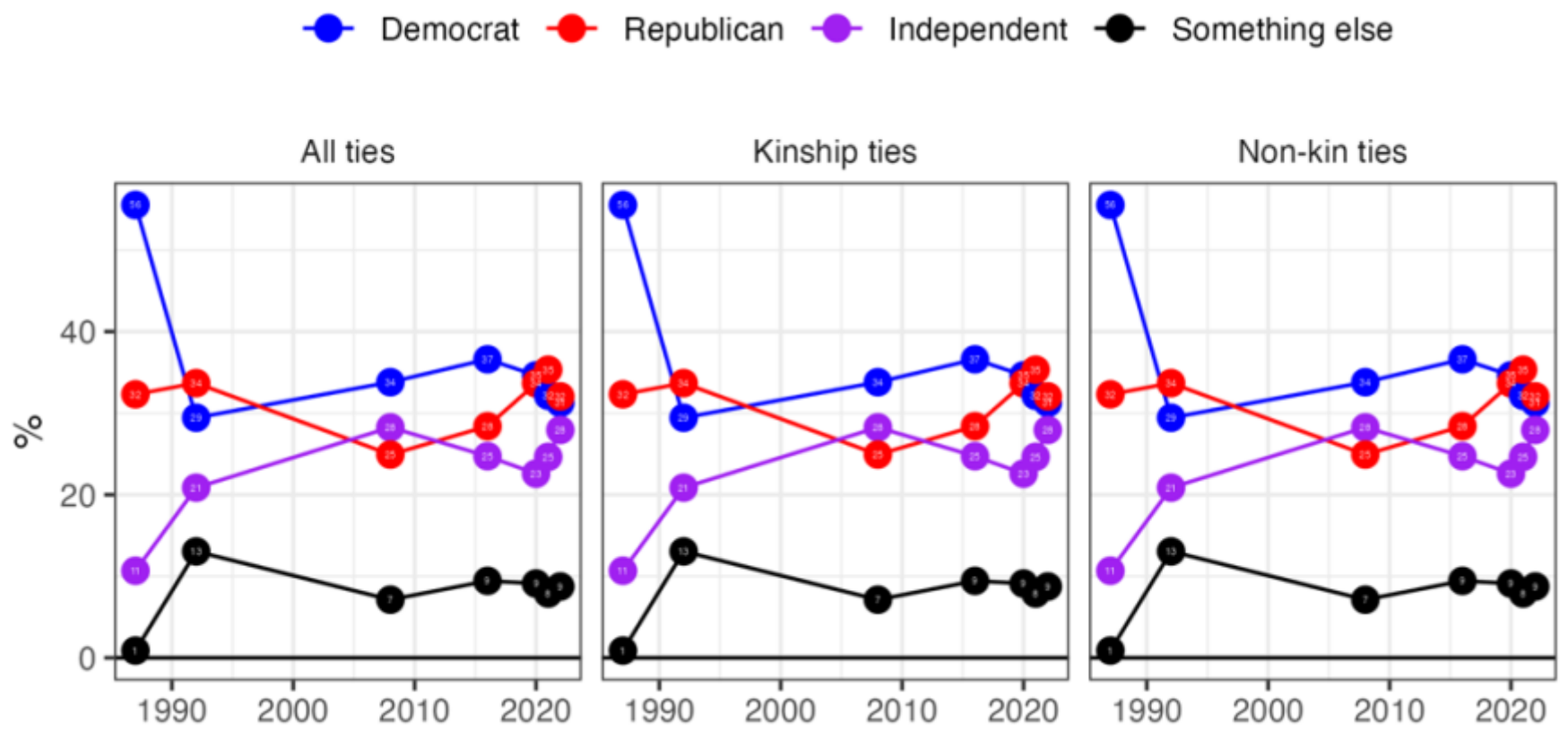

Panel B. The level of absolute homophily: the % same partisanship

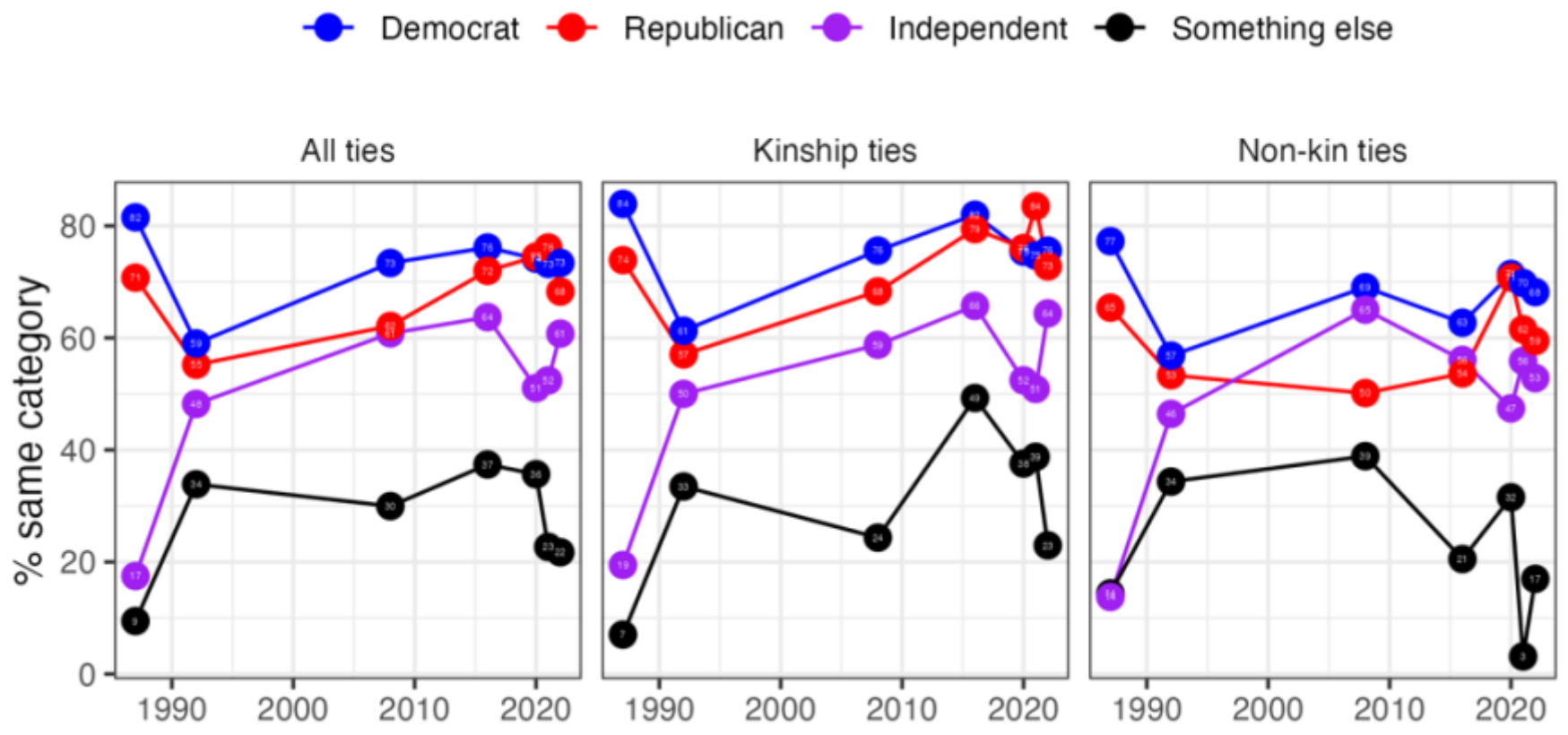

Panel C. The level of choice homophily: Coleman index: from -1 (heterophily) to 1 (homophily)

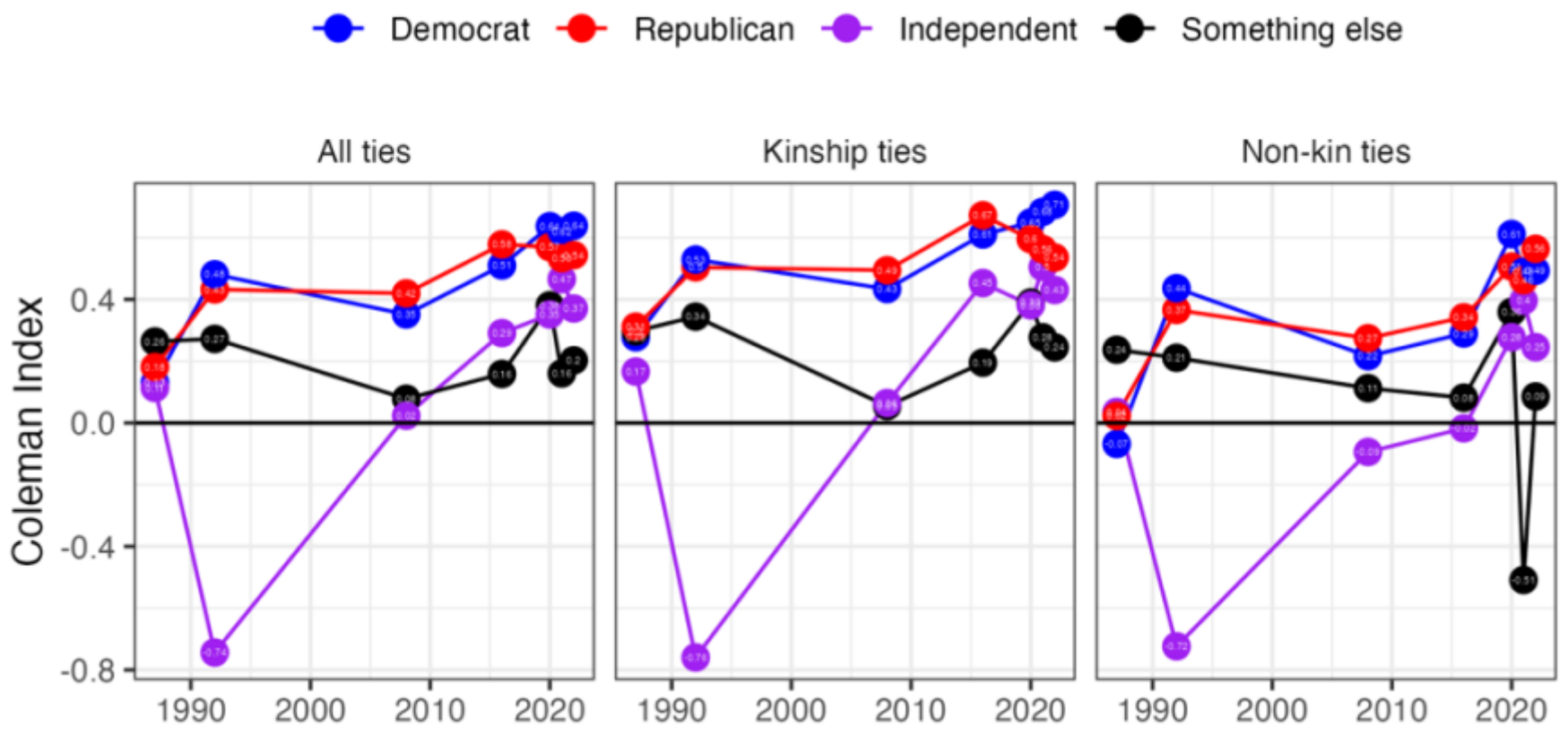

**Fig. S8. The decomposition of political homophily across different partisan groups from 1987 to 2022.** Panel A shows the distribution of each partisanship category over time, which has been used to control for structural opportunities in calculating the Coleman index (Panel C). Panel B shows the level of absolute homophily measured by the proportion of the same partisan ties. Panel C shows the level of choice homophily measured by the Coleman index.

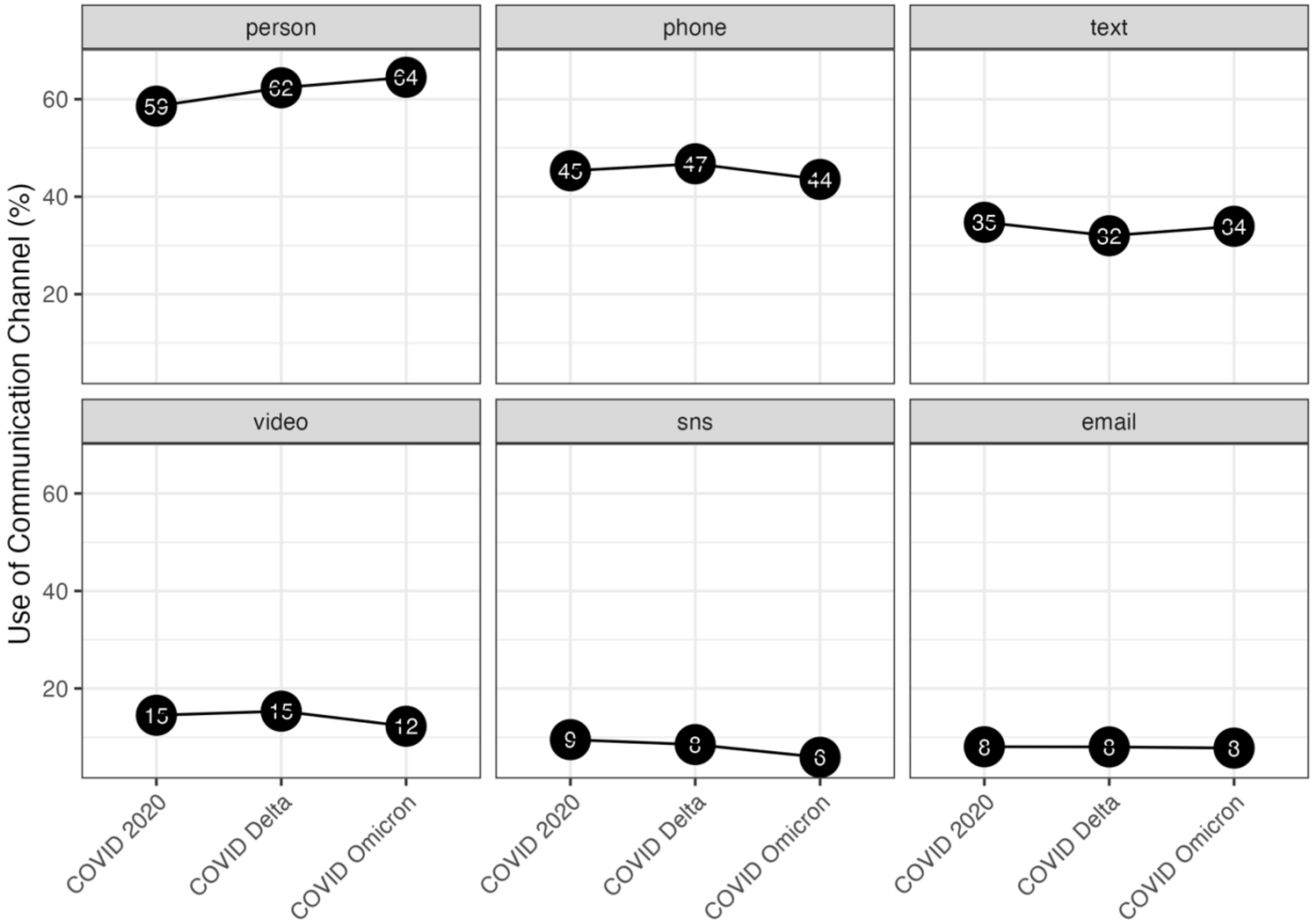

**Fig. S9. The patterns of communication channels during the COVID-19 pandemic.** Weighted proportions with 95% confidence intervals for channel usage with confidants are presented across three phases of the COVID-19 pandemic during the study period.

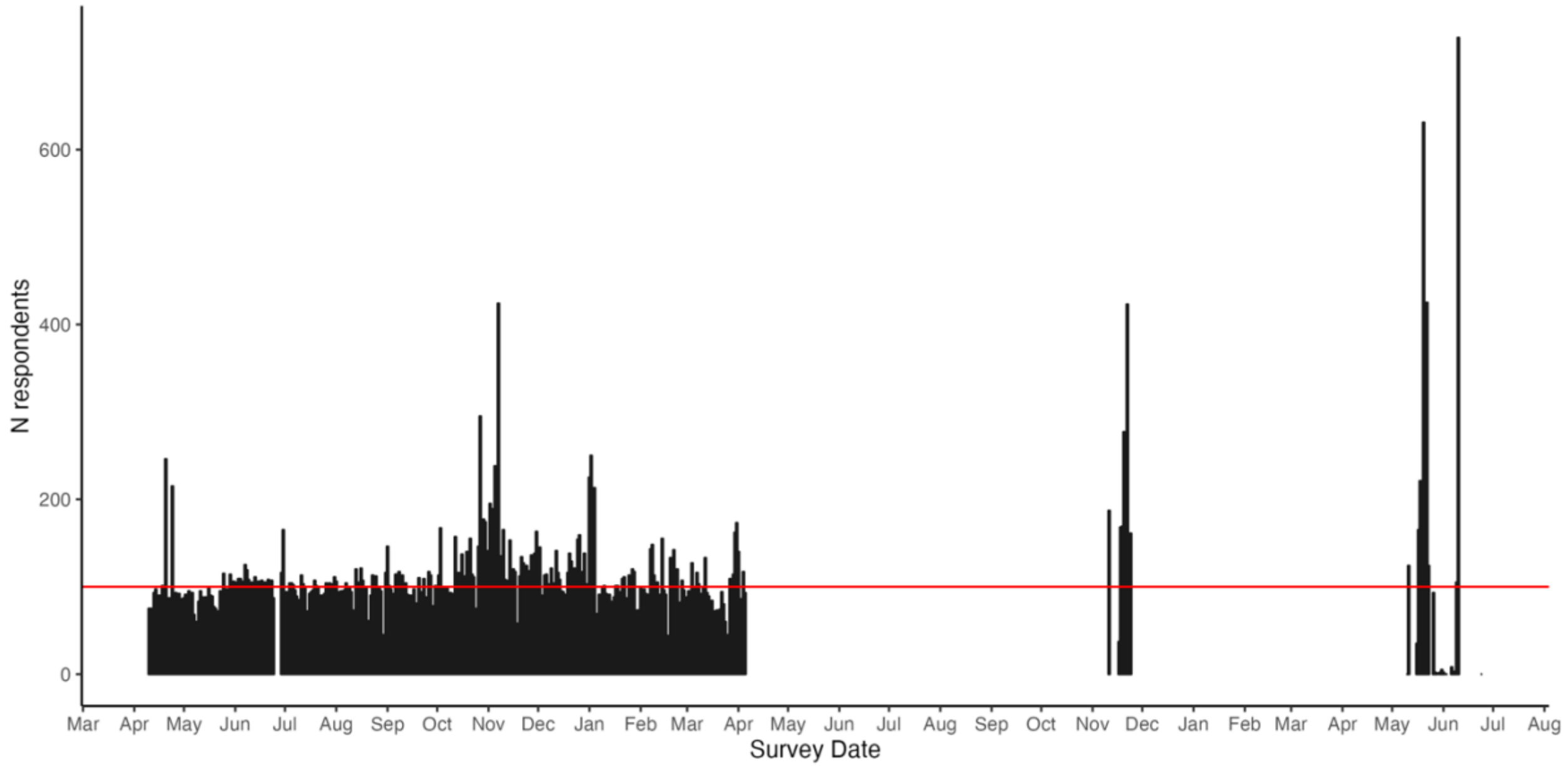

**Fig. S10. The over-time distribution of sample sizes during the entire survey period.** The total number of respondents in the first survey's analytic sample is 36,345 for 357 days from April 10, 2020, to April 5, 2021, which yielded 101.8 competes per day. Here, survey responses for four days (June 25, 26, 27, and 28 in 2020) are missing due to updates on new survey items. We had more completed responses several days in November 2020 and January 2021, though we carefully examined that their demographic characteristics are not significantly different from other cases in the same month. We did not have any daily complete target in the second and third surveys, where survey respondents had been recruited for two weeks in November 2021 (N = 1,776) and for a month in May 2022 (2,912).

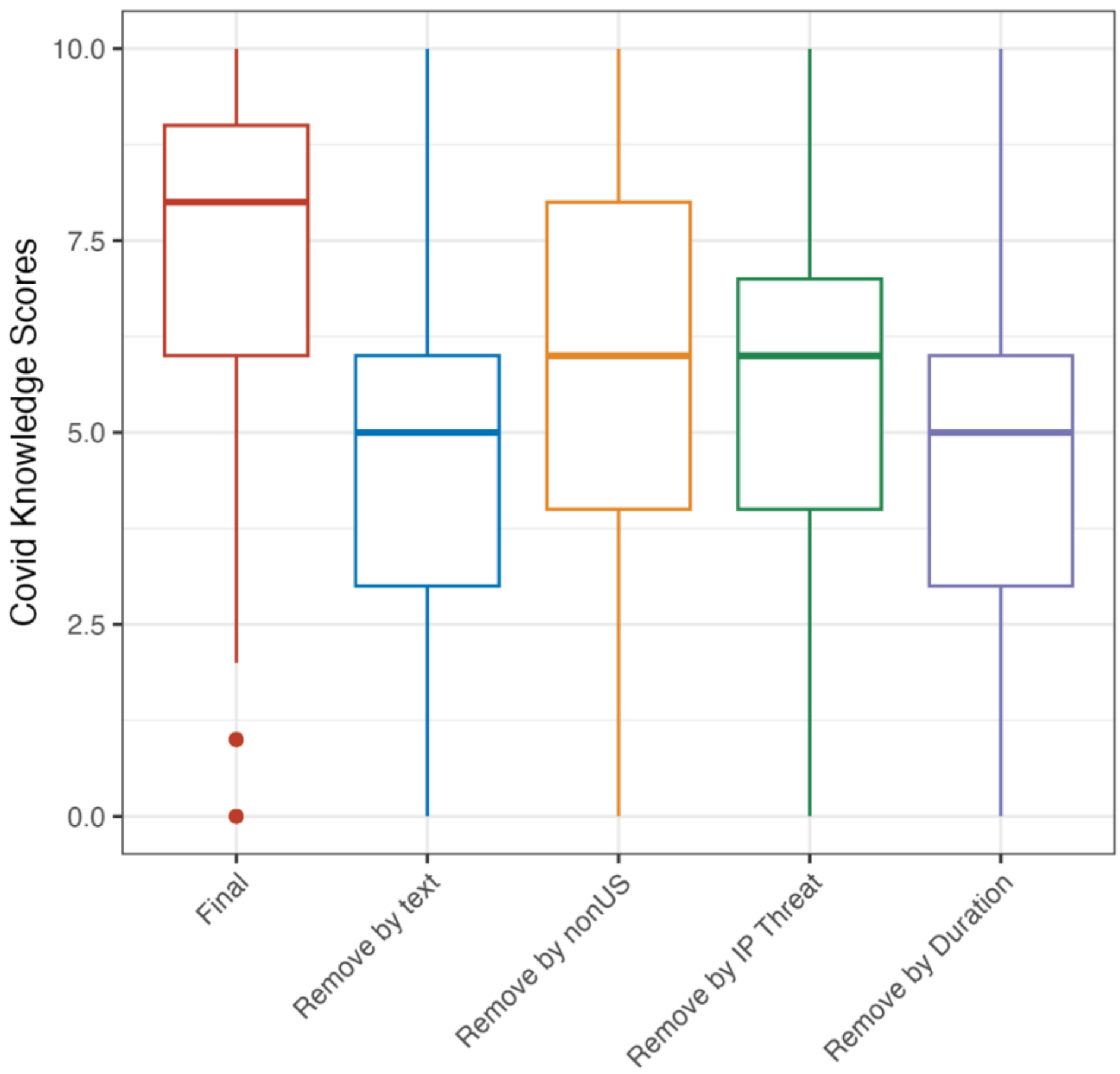

**Fig. S11. COVID-19 knowledge scores across different quality control criteria.** To validate our quality control procedure, we compared the level of knowledge on COVID-19 that we created based on World Health Organization's guideline (https://www.who.int/emergencies/diseases/novel-coronavirus-2019/advice-for-public/myth-busters. Accessed as of April 20, 2020) , "Coronavirus disease (COVID-19) advice for the public: Mythbusters." We asked the following ten questions about COVID-19 and counted the number of True answers for each person: 1. 5G mobile networks spread COVID-19. (False) 2. Antibiotics are effective in preventing and treating the new coronavirus. (False) 3. Drinking alcohol will protect you against COVID-19. (False) 4. Everybody who gets COVID-19 shows symptoms. (False) 5. Eating garlic helps prevent infection with the new coronavirus. (False) 6. The new coronavirus affects only older people. (False) 7. You cannot recover from the coronavirus disease (COVID-19). (False) 8. The new coronavirus can be transmitted through mosquito bites. (False) 9. COVID-19 virus can be transmitted in areas with hot and humid climates. (True) 10. Vaccines against pneumonia cannot protect you against the new coronavirus. (True).

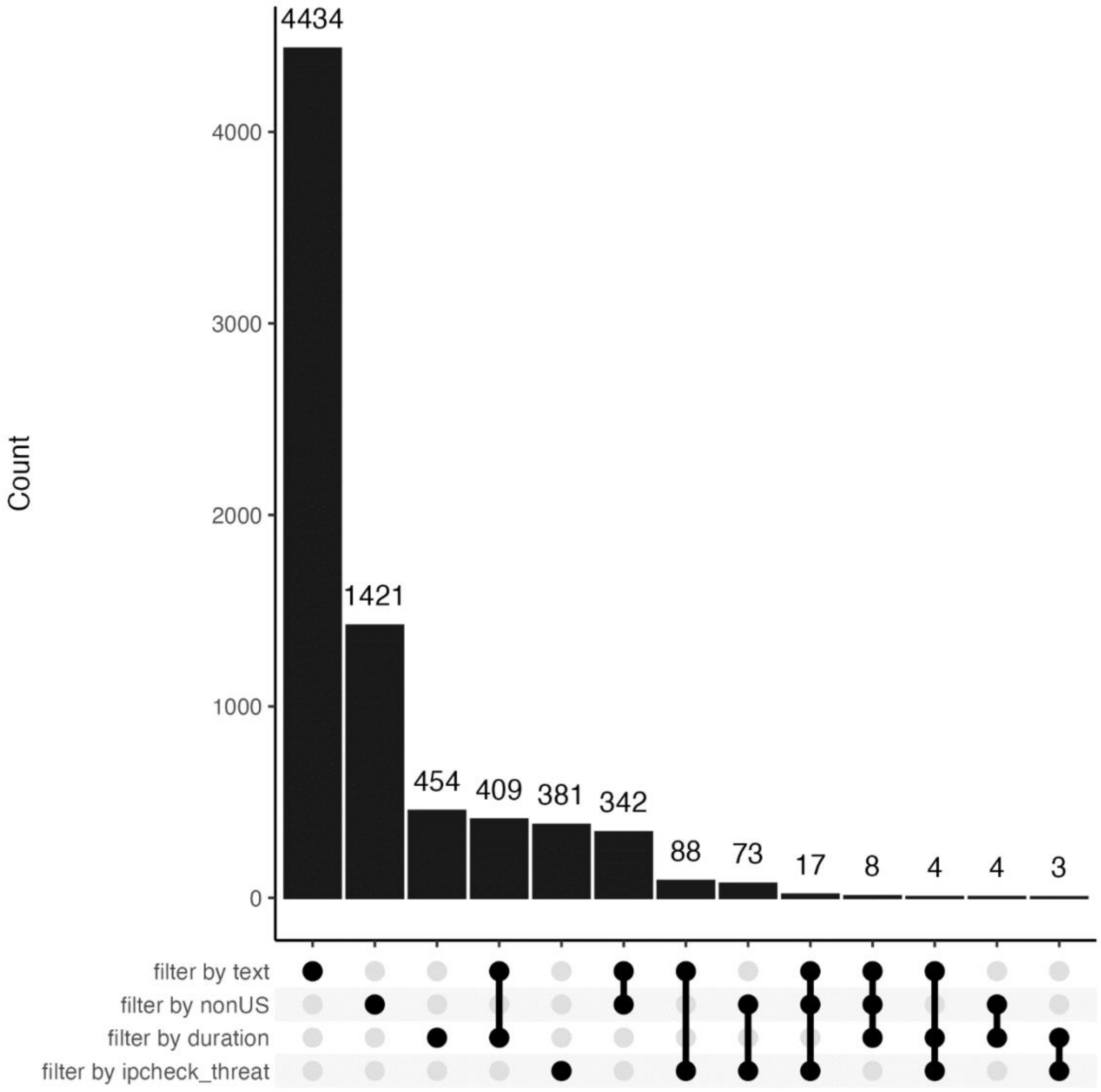

**Fig. S12. The joint distribution of fraudulent responses by different quality control types.**

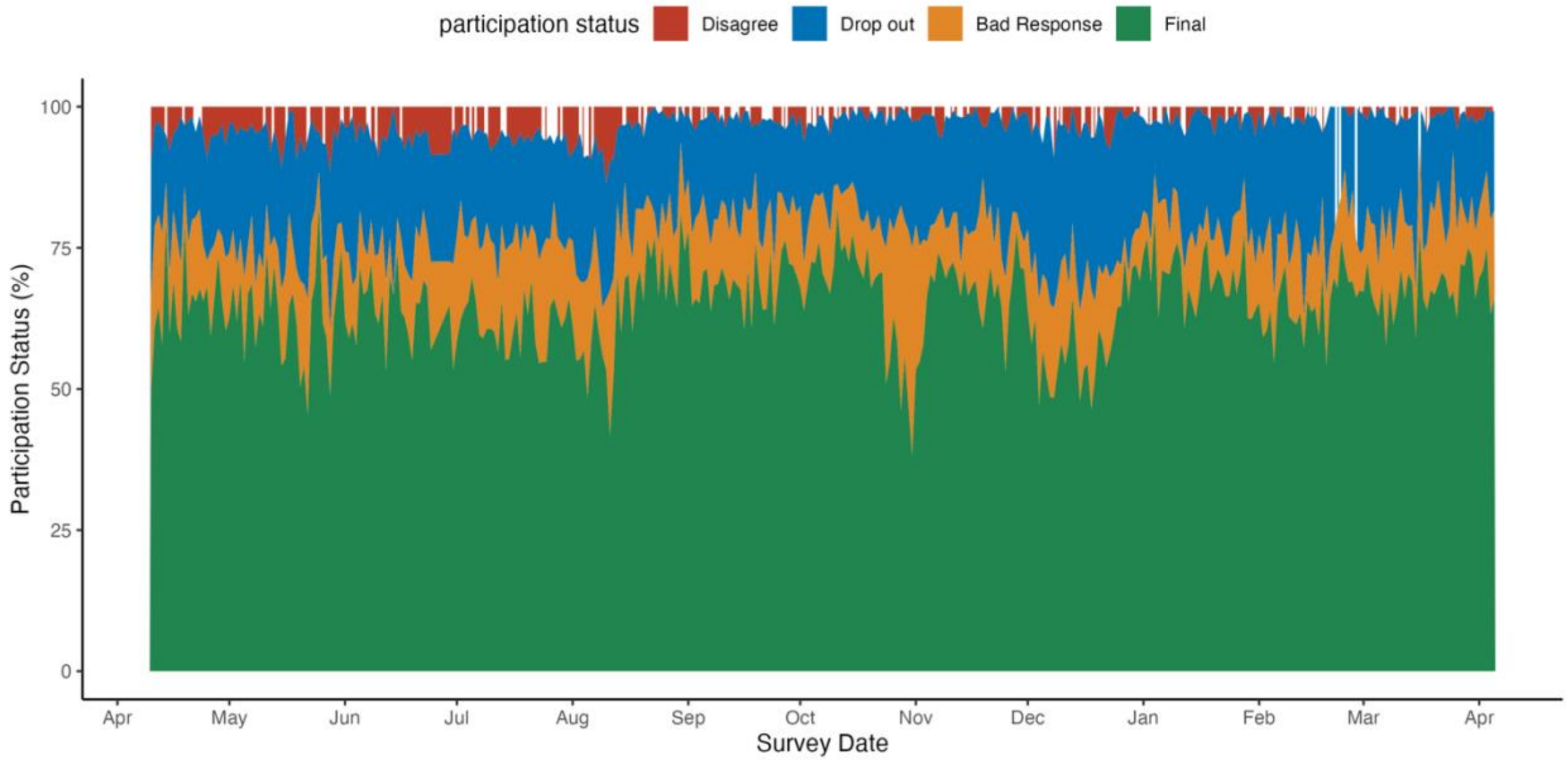

**Fig. S13. The trends of participation status from April 2020 to April 2021.** Among 56,280 individuals who opened the survey link in the entire period in our first survey, 3.2% did not agree to participate, 19.6% dropped out in the middle of the survey, 12.5% were classified as bad/fraudulent responses, and 64.6% consisted of the analytic sample in our first survey. In the second survey, among 2,506 individuals who opened the survey link, 2.7% did not agree to participate, 14.0% dropped out in the middle of the survey, 12.4% were classified as bad/fraudulent responses, and 70.9% consisted of the analytic sample. In the third survey, among 4,222 individuals who opened the survey link, 3.4% did not agree to participate, 20.9% dropped out in the middle of the survey, 6.8% were classified as bad/fraudulent responses, and finally 69.0% consisted of the analytic sample.

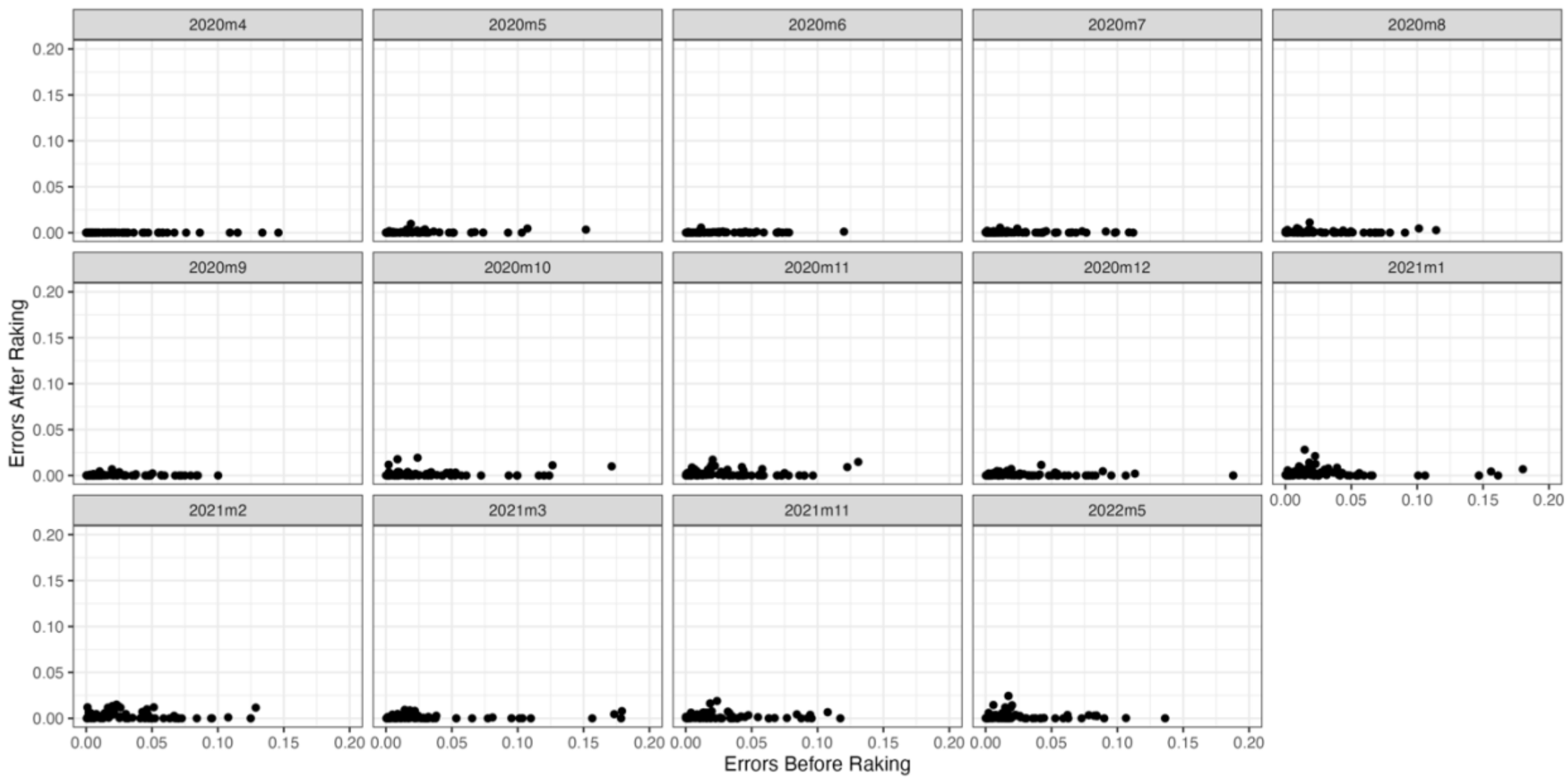

**Fig. S14. Changes in sampling bias for demographic variables before and after raking.** We calculate sampling bias for 10 variables and 98 categories used in the raking procedure by taking the absolute difference in the proportion of each variable from our sample against that from current population survey (CPS) data. Each dot represents the amount of bias for a variable before (x-axis) and after (y-axis). The mean biases across all months before and after raking are 0.023 and 0.001 respectively.

**Table S1. The distribution of network sizes across different surveys.**

| Data set | GSS | GSS | CNES | GSS | PEW | GSS | TESS | TESS | COVID | COVID | COVID |
|---|---|---|---|---|---|---|---|---|---|---|---|
| Year | 1985 | 1987 | 1992 | 2004 | 2008 | 2010 | 2010 | 2016 | 2020/21 | 2021 Nov | 2022 May |
| % of Isolation | 8.1 | 19 | 22.6 | 10.5 | 12.7 | 4.1 | 26.7 | 13.3 | 15.6 | 15.7 | 8.1 |
| Size = 1 | 14.8 | 13.7 | 19.8 | 35.5 | 21.4 | 31.2 | 37.6 | 41.5 | 40.2 | 41.5 | 14.8 |
| Size = 2 | 14.7 | 20.7 | 19.8 | 23.5 | 23.1 | 17 | 17.8 | 24 | 26.9 | 23.9 | 14.7 |
| Size = 3 | 21.6 | 16.6 | 17.1 | 15.7 | 19.6 | 14.9 | 11.5 | 12 | 7.5 | 9.4 | 21.6 |
| Size = 4 | 15.4 | 30 | 9.2 | 8 | 9.4 | 10.4 | 2.9 | 5.1 | 5.5 | 5.8 | 15.4 |
| Size = 5 | 20 | | 6.6 | 6.9 | 7.7 | 7.1 | 2.4 | 1.9 | 1.3 | 1.4 | 20 |
| Size = 6 | 5.4 | | 4.8 | | 6.1 | 5.2 | 1.2 | 2.1 | 3 | 2.3 | 5.4 |
| Size = 6 + | | | | | | 10.1 | | | | | |
| Network Size | | | | | | | | | | | |
| Max | 6 | 4 | 6 | 5 | 6 | 70 | 6 | 6 | 6 | 6 | 6 |
| Mean | 3.03146601 | 2.24886191 | 2.09697266 | 1.95713191 | 2.39037156 | 3.27305228 | 1.3801565 | 1.68442585 | 1.63028437 | 1.61505361 | 3.03146601 |
| Mean (capped at six) | 3.03 | 2.25 | 2.1 | 1.96 | 2.39 | 2.84 | 1.38 | 1.68 | 1.63 | 1.62 | 3.03 |
| Mean (capped at five) | 2.97700029 | 2.24886191 | 2.04853845 | 1.95713191 | 2.32971588 | 2.6325218 | 1.36860747 | 1.66314763 | 1.6001526 | 1.59211264 | 2.97700029 |
| N | 1531 | 1318 | 1426 | 2128 | 1272 | 2061 | 526 | 36345 | 1776 | 2912 | 1531 |
| Survey Period | spring | spring | fall | fall | fall | summer | spring | fall | all seasons | fall | spring |
| Presidential Election | no | no | yes | yes | no | no | no | yes | yes | no | no |
| Survey Mode | FF/tel. | FF/tel. | tel. | FF/tel. | tel. | FF/tel. | internet | internet | internet | internet | internet |

Note. FF: face-to-face, tel: telephone.

**Table S2. Relationship compositions in core discussion networks across different surveys from 1985 to 2022.**

| Dataset | GSS | CNES | GSS | PEW | GSS | TESS | TESS | COVID | COVID | COVID |
|---|---|---|---|---|---|---|---|---|---|---|
| Year | 1985 | 1992 | 2004 | 2008 | 2010 | 2010 | 2016 | 2020/21 | 2021 Nov | 2022 May |
| Parent | 12.3 | | 13.2 | 10.9 | 13.4 | 12.2 | 12.5 | 12.8 | 12.2 | 10.9 |
| Sibling | 9 | | 9.1 | 10.8 | 10.3 | 9.9 | 3.8 | 9.9 | 9.9 | 10.4 |
| Spouse | 14.5 | 15.1 | 19.2 | 17.4 | 17.3 | 22.3 | 49 | 25.6 | 26.7 | 26.3 |
| Child | 9 | | 8.1 | 12.4 | 10.2 | 6.7 | 6.8 | 11.7 | 9.8 | 10.7 |
| Other family | 8.4 | 34.7 | 7.2 | 10.7 | 6.6 | 7 | 3.7 | 7.4 | 6.9 | 6.6 |
| Coworker | 8 | 8.1 | 7.6 | 6.8 | 6 | 3.7 | 5.4 | 3.8 | 3.4 | 3.2 |
| Friend | 27.7 | 33 | 26.6 | 24 | 25.8 | 32.1 | 16.3 | 23.2 | 23.3 | 24.8 |
| Neighbor | 3.1 | 4.7 | 2 | 1.2 | 1.6 | 0.9 | 0.7 | 1.6 | 1.9 | 1.7 |
| Other | 8 | 4.4 | 7.1 | 5.7 | 8.8 | 5.2 | 1.8 | 4 | 5.9 | 5.6 |

Note. To account for the fact that multiple responses are allowed for the GSS and CNES studies, we run 1000 random selections of relationship categories and take the average across 1000 runs in 1985, 1992, 2004, 2008, and 2010. Specifically, we use kin-based random selection: first randomly select one relationship category among kin, and then select one relationship category among other categories, based on the assumption that people would prioritize kin ties over non-kin ties. The mean proportion of each category across 100 simulations is presented here.

**Table S3. Comparison of the distribution of network ties by alters' location and the use of in-person channels between the 2008 PEW survey and our COVID survey.**

| Channel | Location | COVID 2020 | COVID Delta (2021 Nov) | COVID Omicron (2022 May) | PEW-a | PEW-b | PEW-c |
|---|---|---|---|---|---|---|---|
| In-Person | Same household | 32.7 | 32.3 | 31.4 | 21.8 | 21.8 | 21.8 |
| Not in-person | Same household | 3.2 | 4.4 | 2.7 | 0.1 | 0 | 0 |
| In-Person | Different household | 23.9 | 27.8 | 29.1 | 40.4 | 54.7 | 64.5 |
| Not in-person | Different household | 40.1 | 35.5 | 36.8 | 37.7 | 23.4 | 13.6 |

Note. The total number of dyads is 4,259 for the 2008 PEW survey and 71,935 for our COVID survey (64,220 for COVID 2020, 2,958 for COVID Delta wave, and 4,757 for COVID Omicron wave). Both the 2008 PEW survey and our COVID survey collect information on whether their alters lived in the same household similarly. In contrast to our COVID survey that asks whether respondents used an in-person channel the last time they talked to their alters, the 2008 PEW survey asks how often respondents had a face-to-face conversation with their alters over the last six months with the response options: several times a day (=1), once a day (=2), several times a week (=3), once a week (=4), once a month (=5), less often (=6), and never (=7). Here, we show results from comparisons with different thresholds to estimate the use of in-person channels in the last conversation in the PEW 2008 survey: a = 1 to 3, b = 1 to 4, c = 1 to 5. We discuss version b (PEW-b) in the manuscript.

**Table S4. Results from logistic regression models predicting political homophily across kin and non-kin ties across three phases of the pandemic.**

| Period | COVID 2020 | COVID 2020 | All | All |
|---|---|---|---|---|
| Sample | Kin | Nonkin | Kin | Nonkin |
| Model | Model1 | Model2 | Model3 | Model4 |
| inperson only | 0.0571 | -0.185* | 0 | 0 |
| | (0.0572) | (0.0874) | (.) | (.) |
| remote and inperson | 0 | 0 | -0.0688 | 0.179* |
| | (.) | (.) | (0.0569) | (0.0869) |
| remote only | -0.0809 | 0.0129 | -0.147* | 0.199* |
| | (0.0639) | (0.0721) | (0.0636) | (0.0784) |
| COVID 2020 (p1) | | | 0 | 0 |
| | | | (.) | (.) |
| COVID 2021 (p2) | | | 0.0620 | 0.0727 |
| | | | (0.166) | (0.279) |
| COVID 2022 (p3) | | | 0.0418 | -0.363 |
| | | | (0.168) | (0.284) |
| p1 X inperson only | | | 0 | 0 |
| | | | (.) | (.) |
| p1 X remote and inperson | | | 0 | 0 |
| | | | (.) | (.) |
| p1 X remote only | | | 0 | 0 |
| | | | (.) | (.) |
| p2 X inperson only | | | 0 | 0 |
| | | | (.) | (.) |
| p2 X remote and inperson | | | 0.194 | 0.109 |
| | | | (0.279) | (0.384) |
| p2 X remote only | | | -0.0941 | -0.276 |
| | | | (0.234) | (0.357) |
| p3 X inperson only | | | 0 | 0 |
| | | | (.) | (.) |
| p3 X remote and inperson | | | -0.186 | 0.444 |
| | | | (0.230) | (0.364) |
| p3 X remote only | | | 0.000712 | -0.147 |
| | | | (0.217) | (0.332) |
| Male | 0.00857 | 0.141* | -0.0286 | 0.120+ |
| | (0.0513) | (0.0664) | (0.0489) | (0.0636) |
| Female | 0 | 0 | 0 | 0 |
| | (.) | (.) | (.) | (.) |
| Race = White | 0 | 0 | 0 | 0 |
| | (.) | (.) | (.) | (.) |
| Black | 0.372*** | 0.551*** | 0.386*** | 0.525*** |
| | (0.0918) | (0.118) | (0.0878) | (0.114) |
| Asian | -0.169 | -0.291+ | -0.164 | -0.326* |
| | (0.113) | (0.160) | (0.107) | (0.150) |
| Other | -0.123 | -0.318+ | -0.158 | -0.339* |
| | (0.161) | (0.171) | (0.149) | (0.163) |
| Hispanic | 0.00854 | -0.197+ | -0.0229 | -0.195+ |
| | (0.0786) | (0.105) | (0.0751) | (0.0998) |
| Age -29 | -0.304*** | 0.0563 | -0.347*** | 0.0427 |
| | (0.0851) | (0.122) | (0.0799) | (0.115) |
| Age 30-39 | 0 | 0 | 0 | 0 |
| | (.) | (.) | (.) | (.) |
| Age 40-49 | 0.238** | 0.0939 | 0.160* | 0.0521 |
| | (0.0819) | (0.0974) | (0.0785) | (0.0934) |
| Age 50-59 | 0.301*** | -0.0892 | 0.213** | -0.109 |
| | (0.0867) | (0.106) | (0.0817) | (0.102) |
| Age 60-69 | 0.290** | 0.158 | 0.258** | 0.0958 |
| | (0.0905) | (0.119) | (0.0874) | (0.114) |
| Age 70+ | 0.565*** | 0.216 | 0.487*** | 0.132 |
| | (0.114) | (0.144) | (0.108) | (0.137) |
| Less than high school | -0.0879 | 0.0386 | -0.0945 | 0.0632 |
| | (0.118) | (0.170) | (0.113) | (0.162) |
| HS | 0 | 0 | 0 | 0 |
| | (.) | (.) | (.) | (.) |
| Some college | -0.0367 | 0.123 | -0.0412 | 0.0708 |
| | (0.0631) | (0.0903) | (0.0601) | (0.0863) |
| BS+ | 0.123+ | 0.327*** | 0.0885 | 0.296*** |
| | (0.0710) | (0.0912) | (0.0674) | (0.0868) |
| Household Income less than 10K | 0.0665 | -0.0858 | -0.00143 | -0.0366 |
| | (0.121) | (0.148) | (0.114) | (0.143) |
| 10K- | -0.00206 | -0.275* | -0.0215 | -0.278* |
| | (0.105) | (0.137) | (0.0996) | (0.130) |
| 20K- | 0.0680 | -0.0252 | -0.00182 | -0.0730 |
| | (0.0875) | (0.125) | (0.0831) | (0.118) |
| 30K- | 0.0351 | -0.177 | 0.0224 | -0.0619 |
| | (0.0904) | (0.114) | (0.0862) | (0.108) |
| 40K- | 0.0819 | -0.146 | 0.0612 | -0.0534 |
| | (0.0885) | (0.121) | (0.0851) | (0.118) |
| 50K- | 0.107 | 0.00409 | 0.130 | 0.00129 |
| | (0.0879) | (0.117) | (0.0828) | (0.113) |
| 60K- | 0 | 0 | 0 | 0 |
| | (.) | (.) | (.) | (.) |
| 100K- | -0.0368 | 0.106 | -0.0488 | 0.0575 |
| | (0.0798) | (0.101) | (0.0772) | (0.0972) |
| 150K- | 0.252* | 0.0977 | 0.272** | 0.106 |
| | (0.0989) | (0.113) | (0.0937) | (0.108) |
| Married | 0 | 0 | 0 | 0 |
| | (.) | (.) | (.) | (.) |
| Widowed | -0.186 | -0.205 | -0.211+ | -0.261+ |
| | (0.115) | (0.147) | (0.110) | (0.141) |
| Divorced | -0.282*** | -0.220* | -0.251** | -0.245* |
| | (0.0844) | (0.103) | (0.0804) | (0.0970) |
| Separated | -0.566*** | -0.373* | -0.484** | -0.312+ |
| | (0.158) | (0.187) | (0.151) | (0.177) |
| Never married | -0.0671 | -0.158 | -0.0463 | -0.152 |
| | (0.0721) | (0.105) | (0.0682) | (0.0997) |
| DK | -0.0191 | -0.159 | -0.0202 | -0.198 |
| | (0.238) | (0.308) | (0.229) | (0.295) |
| Household size = 1 | 0 | 0 | 0 | 0 |
| | (.) | (.) | (.) | (.) |
| 2 | -0.0985 | -0.0103 | -0.128+ | -0.0703 |
| | (0.0787) | (0.101) | (0.0752) | (0.0959) |
| 3 | -0.0537 | -0.106 | -0.0545 | -0.0799 |
| | (0.0904) | (0.113) | (0.0860) | (0.107) |
| 4 | -0.0230 | -0.0854 | -0.0481 | -0.0625 |
| | (0.0972) | (0.120) | (0.0924) | (0.116) |
| 5 | -0.114 | 0.0112 | -0.177 | 0.0269 |
| | (0.115) | (0.154) | (0.109) | (0.149) |
| more than 5 | -0.149 | -0.202 | -0.137 | -0.208 |
| | (0.127) | (0.161) | (0.120) | (0.153) |
| Republican | 0 | 0 | 0 | 0 |
| | (.) | (.) | (.) | (.) |
| Democrat | 0.165** | 0.238*** | 0.207*** | 0.201** |
| | (0.0575) | (0.0690) | (0.0551) | (0.0665) |
| Independent | -1.058*** | -1.003*** | -1.003*** | -0.992*** |
| | (0.0625) | (0.0878) | (0.0597) | (0.0838) |
| Something else | -1.940*** | -1.743*** | -1.952*** | -1.861*** |
| | (0.128) | (0.224) | (0.121) | (0.216) |
| Not working now | -0.339 | 0.132 | -0.274 | -0.0124 |
| | (0.207) | (0.327) | (0.195) | (0.304) |
| Other | -0.173* | 0.106 | -0.156* | 0.0861 |
| | (0.0757) | (0.107) | (0.0719) | (0.102) |
| Retired | -0.150+ | 0.0524 | -0.153+ | 0.0949 |
| | (0.0833) | (0.110) | (0.0792) | (0.104) |
| Unable to work | -0.452** | -0.249 | -0.425** | -0.269 |
| | (0.142) | (0.290) | (0.140) | (0.273) |
| Unemployed | -0.312*** | -0.0247 | -0.311*** | -0.0708 |
| | (0.0901) | (0.115) | (0.0876) | (0.112) |
| Working now | 0 | 0 | 0 | 0 |
| | (.) | (.) | (.) | (.) |
| metro | -0.0152 | 0.0973 | -0.0219 | 0.106 |
| | (0.0635) | (0.0906) | (0.0610) | (0.0874) |
| Same household | 0 | 0 | 0 | 0 |
| | (.) | (.) | (.) | (.) |
| Same neighborhood | -0.320*** | -0.325** | -0.352*** | -0.368*** |
| | (0.0699) | (0.111) | (0.0664) | (0.108) |
| Same state | -0.444*** | -0.445*** | -0.443*** | -0.467*** |
| | (0.0635) | (0.111) | (0.0611) | (0.107) |
| Somewhere else | -0.571*** | -0.691*** | -0.564*** | -0.717*** |
| | (0.0757) | (0.124) | (0.0716) | (0.119) |
| Constant | 0.856*** | 0.400+ | 1.010*** | 0.332 |
| | (0.146) | (0.230) | (0.140) | (0.220) |
| Observations | 40912 | 22143 | 45760 | 24787 |

Note. Standard errors are in parenthesis (+ p < 0.1, * p < 0.05, ** p < 0.01, *** p < 0.005).

**Table S5. Demographic comparison between the fraudulent sample and the final sample.**

| | filtered | final | p |
|---|---|---|---|
| n | 7638 | 41033 | |
| sex = Female (%) | 2710 (35.5) | 24800 (60.4) | <0.001 |
| age (mean (SD)) | 33.94 (11.11) | 45.63 (16.90) | <0.001 |
| race (%) | | | <0.001 |
| White | 4895 (64.1) | 31196 (76.0) | |
| Black | 1325 (17.3) | 3976 ( 9.7) | |
| Asian | 468 ( 6.1) | 1547 ( 3.8) | |
| Other | 261 ( 3.4) | 1375 ( 3.4) | |
| Hispanic | 689 ( 9.0) | 2939 ( 7.2) | |
| partyid4 (%) | | | <0.001 |
| Republican | 3300 (43.2) | 14376 (35.0) | |
| Democrat | 2578 (33.8) | 15935 (38.8) | |
| Independent | 1238 (16.2) | 8106 (19.8) | |
| Something else | 522 ( 6.8) | 2616 ( 6.4) | |
| educ4 (%) | | | <0.001 |
| Less than high school | 394 ( 5.2) | 1516 ( 3.7) | |
| HS | 1472 (19.3) | 9497 (23.1) | |
| Some college | 1565 (20.5) | 13552 (33.0) | |
| BS+ | 4207 (55.1) | 16468 (40.1) | |
| marital_status (%) | | | <0.001 |
| Married | 4389 (59.4) | 20464 (50.7) | |
| Widowed | 270 ( 3.7) | 2120 ( 5.2) | |
| Divorced | 373 ( 5.1) | 4965 (12.3) | |
| Separated | 280 ( 3.8) | 961 ( 2.4) | |
| Never married | 2072 (28.1) | 11886 (29.4) | |
| empstat (%) | | | <0.001 |
| Not working now | 242 ( 3.2) | 702 ( 1.7) | |
| Other | 785 (10.3) | 4795 (11.7) | |
| Retired | 285 ( 3.7) | 7409 (18.1) | |
| Unable to work | 151 ( 2.0) | 2646 ( 6.4) | |
| Unemployed | 802 (10.5) | 4666 (11.4) | |
| Working now | 5373 (70.3) | 20815 (50.7) | |
| census_division (%) | | | <0.001 |
| East North Central | 826 (10.9) | 5761 (14.0) | |
| East South Central | 316 ( 4.2) | 2547 ( 6.2) | |
| Mid-Atlantic | 1730 (22.9) | 6577 (16.0) | |
| Mountain | 445 ( 5.9) | 3295 ( 8.0) | |
| New England | 258 ( 3.4) | 2189 ( 5.3) | |
| Pacific | 1244 (16.5) | 5091 (12.4) | |
| South Atlantic | 1731 (22.9) | 8760 (21.4) | |
| West North Central | 318 ( 4.2) | 2581 ( 6.3) | |
| West South Central | 691 ( 9.1) | 4227 (10.3) | |
| metro = 1 (%) | 6153 (80.6) | 33979 (82.8) | <0.001 |
| income (%) | | | <0.001 |
| Less than $10,000 | 905 (11.8) | 3685 ( 9.0) | |
| $10,000 to $19,999 | 629 ( 8.2) | 3709 ( 9.0) | |
| $20,000 to $29,999 | 732 ( 9.6) | 4651 (11.3) | |
| $30,000 to $39,999 | 520 ( 6.8) | 4158 (10.1) | |
| $40,000 to $49,999 | 421 ( 5.5) | 3410 ( 8.3) | |
| $50,000 to $59,999 | 512 ( 6.7) | 3488 ( 8.5) | |
| $60,000 to $69,999 | 310 ( 4.1) | 2347 ( 5.7) | |
| $70,000 to $79,999 | 487 ( 6.4) | 2630 ( 6.4) | |
| $80,000 to $89,999 | 215 ( 2.8) | 1495 ( 3.6) | |
| $90,000 to $99,999 | 432 ( 5.7) | 1914 ( 4.7) | |
| $100,000 to 149,999 | 1285 (16.8) | 4981 (12.1) | |
| $150,000 or more | 1078 (14.1) | 3758 ( 9.2) | |
| Prefer not to tell | 112 ( 1.5) | 807 ( 2.0) | |
| household_size (%) | | | <0.001 |
| 1 | 1652 (21.6) | 8360 (20.4) | |
| 2 | 1488 (19.5) | 13280 (32.4) | |
| 3 | 1708 (22.4) | 7591 (18.5) | |
| 4 | 1850 (24.2) | 7284 (17.8) | |
| 5 | 577 ( 7.6) | 2799 ( 6.8) | |
| more than 5 | 363 ( 4.8) | 1719 ( 4.2) | |

Note. P-value is based on the Chi-square tests.

**Table S6. Summary of sample characteristics before and after raking and the target statistics from CPS data in March 2021.**

| Sample | unweighted | weighted | CPS (target) | Diff unweighted | Diff weighted |
|---|---|---|---|---|---|
| N | 3302 | 3302 | 254736187.3 | | |
| Female | 1893.0 (57.3) | 1716.8 (52.0) | 131613078.6 (51.7) | -5.6 | -0.3 |
| Age Group | | | | | |
| Age -29 | 303.0 ( 9.2) | 617.3 (18.7) | 54870055.8 (21.5) | 12.3 | 2.8 |
| Age 30-39 | 884.0 (26.8) | 541.8 (16.4) | 43692842.2 (17.2) | -9.6 | 0.8 |
| Age 40-49 | 541.0 (16.4) | 504.8 (15.3) | 39228128.3 (15.4) | -1 | 0.1 |
| Age 50-59 | 412.0 (12.5) | 536.2 (16.2) | 41007634.0 (16.1) | 3.6 | -0.1 |
| Age 60-69 | 653.0 (19.8) | 546.2 (16.6) | 38754387.6 (15.2) | -4.6 | -1.4 |
| Age 70+ | 507.0 (15.4) | 554.1 (16.8) | 37183139.3 (14.6) | -0.8 | -2.2 |
| Race | | | | | |
| White | 2785.0 (84.3) | 2195.7 (66.5) | 158466542.8 (62.2) | -22.1 | -4.3 |
| Black | 225.0 ( 6.8) | 326.7 ( 9.9) | 30803576.6 (12.1) | 5.3 | 2.2 |
| Asian | 69.0 ( 2.1) | 186.8 ( 5.7) | 15753631.6 ( 6.2) | 4.1 | 0.5 |
| Other | 78.0 ( 2.4) | 84.7 ( 2.6) | 6424515.7 ( 2.5) | 0.1 | -0.1 |
| Hispanic | 145.0 ( 4.4) | 508.1 (15.4) | 43287920.6 (17.0) | 12.6 | 1.6 |
| Education | | | | | |
| less than high school | 643.0 (19.5) | 977.8 (29.6) | 75414390.1 (29.6) | 10.1 | 0 |
| High school | 86.0 ( 2.6) | 301.9 ( 9.1) | 24149374.4 ( 9.5) | 6.9 | 0.4 |
| some college | 939.0 (28.4) | 422.0 (12.8) | 32035571.8 (12.6) | -15.8 | -0.2 |
| college | 698.0 (21.1) | 720.5 (21.8) | 55744049.8 (21.9) | 0.8 | 0.1 |
| master | 936.0 (28.3) | 879.8 (26.6) | 67392801.2 (26.5) | -1.8 | -0.1 |
| Working Status | | | | | |
| Working now | 1484.0 (44.9) | 1578.4 (47.8) | 125668850.2 (49.3) | 4.4 | 1.5 |
| Self-employed | 282.0 ( 8.5) | 179.1 ( 5.4) | 12777064.7 ( 5.0) | -3.5 | -0.4 |
| Temporarily laid off | 10.0 ( 0.3) | 8.3 ( 0.3) | 634011.3 ( 0.2) | -0.1 | -0.1 |
| Retired | 809.0 (24.5) | 720.7 (21.8) | 49420969.4 (19.4) | -5.1 | -2.4 |
| Student | 74.0 ( 2.2) | 1.3 ( 0.0) | 89974.5 ( 0.0) | -2.2 | 0 |
| Maternity leave | 3.0 ( 0.1) | 4.2 ( 0.1) | 319553.8 ( 0.1) | 0 | 0 |
| Illness/Sick leave | 10.0 ( 0.3) | 14.5 ( 0.4) | 1118538.0 ( 0.4) | 0.1 | 0 |
| Disabled | 200.0 ( 6.1) | 102.1 ( 3.1) | 7693509.3 ( 3.0) | -3.1 | -0.1 |
| Other | 220.0 ( 6.7) | 475.4 (14.4) | 38996757.2 (15.3) | 8.6 | 0.9 |
| Unemployed | 210.0 ( 6.4) | 217.9 ( 6.6) | 18016959.0 ( 7.1) | 0.7 | 0.5 |
| State | | | | | |
| Alabama | 44.0 ( 1.3) | 49.8 ( 1.5) | 3827913.6 ( 1.5) | 0.2 | 0 |
| Alaska | 4.0 ( 0.1) | 15.2 ( 0.5) | 532716.4 ( 0.2) | 0.1 | -0.3 |
| Arizona | 94.0 ( 2.8) | 61.1 ( 1.9) | 5367806.8 ( 2.1) | -0.7 | 0.2 |
| Arkansas | 32.0 ( 1.0) | 40.9 ( 1.2) | 2320775.6 ( 0.9) | -0.1 | -0.3 |
| California | 280.0 ( 8.5) | 415.3 (12.6) | 30580518.4 (12.0) | 3.5 | -0.6 |
| Colorado | 40.0 ( 1.2) | 54.1 ( 1.6) | 4580211.9 ( 1.8) | 0.6 | 0.2 |
| Connecticut | 37.0 ( 1.1) | 11.2 ( 0.3) | 2852561.5 ( 1.1) | 0 | 0.8 |
| Delaware | 14.0 ( 0.4) | 6.0 ( 0.2) | 792014.8 ( 0.3) | -0.1 | 0.1 |
| District of Columbia | 8.0 ( 0.2) | 40.9 ( 1.2) | 578767.7 ( 0.2) | 0 | -1 |
| Florida | 239.0 ( 7.2) | 253.8 ( 7.7) | 17524281.0 ( 6.9) | -0.3 | -0.8 |
| Georgia | 116.0 ( 3.5) | 38.5 ( 1.2) | 8212982.6 ( 3.2) | -0.3 | 2 |
| Hawaii | 9.0 ( 0.3) | 37.0 ( 1.1) | 1057219.4 ( 0.4) | 0.1 | -0.7 |
| Idaho | 21.0 ( 0.6) | 91.4 ( 2.8) | 1401922.9 ( 0.6) | 0 | -2.2 |
| Illinois | 125.0 ( 3.8) | 96.0 ( 2.9) | 9754734.5 ( 3.8) | 0 | 0.9 |
| Indiana | 46.0 ( 1.4) | 37.0 ( 1.1) | 5207777.0 ( 2.0) | 0.6 | 0.9 |
| Iowa | 33.0 ( 1.0) | 44.4 ( 1.3) | 2444730.7 ( 1.0) | 0 | -0.3 |
| Kansas | 24.0 ( 0.7) | 27.0 ( 0.8) | 2207213.0 ( 0.9) | 0.2 | 0.1 |
| Kentucky | 52.0 ( 1.6) | 25.0 ( 0.8) | 3465954.1 ( 1.4) | -0.2 | 0.6 |
| Louisiana | 38.0 ( 1.2) | 90.7 ( 2.7) | 3505606.8 ( 1.4) | 0.2 | -1.3 |
| Maine | 6.0 ( 0.2) | 25.7 ( 0.8) | 1112656.3 ( 0.4) | 0.2 | -0.4 |
| Maryland | 59.0 ( 1.8) | 54.6 ( 1.7) | 4700342.5 ( 1.8) | 0 | 0.1 |
| Massachusetts | 71.0 ( 2.2) | 84.9 ( 2.6) | 5361482.8 ( 2.1) | -0.1 | -0.5 |
| Michigan | 118.0 ( 3.6) | 59.2 ( 1.8) | 7867173.0 ( 3.1) | -0.5 | 1.3 |
| Minnesota | 48.0 ( 1.5) | 52.1 ( 1.6) | 4392346.5 ( 1.7) | 0.2 | 0.1 |
| Mississippi | 18.0 ( 0.5) | 22.6 ( 0.7) | 2250410.5 ( 0.9) | 0.4 | 0.2 |
| Missouri | 55.0 ( 1.7) | 18.1 ( 0.5) | 4769769.8 ( 1.9) | 0.2 | 1.4 |
| Montana | 12.0 ( 0.4) | 36.4 ( 1.1) | 854044.4 ( 0.3) | -0.1 | -0.8 |
| Nebraska | 16.0 ( 0.5) | 21.1 ( 0.6) | 1451867.1 ( 0.6) | 0.1 | 0 |
| Nevada | 34.0 ( 1.0) | 71.4 ( 2.2) | 2461105.2 ( 1.0) | 0 | -1.2 |
| New Hampshire | 17.0 ( 0.5) | 23.3 ( 0.7) | 1117944.8 ( 0.4) | -0.1 | -0.3 |
| New Jersey | 95.0 ( 2.9) | 84.8 ( 2.6) | 6964543.2 ( 2.7) | -0.2 | 0.1 |
| New Mexico | 11.0 ( 0.3) | 56.2 ( 1.7) | 1630415.7 ( 0.6) | 0.3 | -1.1 |
| New York | 405.0 (12.3) | 141.0 ( 4.3) | 15243413.0 ( 6.0) | -6.3 | 1.7 |
| North Carolina | 88.0 ( 2.7) | 97.7 ( 3.0) | 8319817.9 ( 3.3) | 0.6 | 0.3 |
| North Dakota | 4.0 ( 0.1) | 20.4 ( 0.6) | 577208.2 ( 0.2) | 0.1 | -0.4 |
| Ohio | 128.0 ( 3.9) | 95.4 ( 2.9) | 9158296.4 ( 3.6) | -0.3 | 0.7 |
| Oklahoma | 36.0 ( 1.1) | 42.6 ( 1.3) | 1962390.2 ( 0.8) | -0.3 | -0.5 |
| Oregon | 39.0 ( 1.2) | 53.0 ( 1.6) | 3418942.9 ( 1.3) | 0.1 | -0.3 |
| Pennsylvania | 144.0 ( 4.4) | 69.6 ( 2.1) | 10160813.9 ( 4.0) | -0.4 | 1.9 |
| Rhode Island | 15.0 ( 0.5) | 31.1 ( 0.9) | 851256.0 ( 0.3) | -0.2 | -0.6 |
| South Carolina | 57.0 ( 1.7) | 94.6 ( 2.9) | 4107357.0 ( 1.6) | -0.1 | -1.3 |
| South Dakota | 13.0 ( 0.4) | 65.2 ( 2.0) | 671772.7 ( 0.3) | -0.1 | -1.7 |
| Tennessee | 67.0 ( 2.0) | 104.9 ( 3.2) | 5399537.5 ( 2.1) | 0.1 | -1.1 |
| Texas | 210.0 ( 6.4) | 161.8 ( 4.9) | 22047521.8 ( 8.7) | 2.3 | 3.8 |
| Utah | 26.0 ( 0.8) | 54.4 ( 1.6) | 2044110.4 ( 0.8) | 0 | -0.8 |
| Vermont | 5.0 ( 0.2) | 16.8 ( 0.5) | 510680.9 ( 0.2) | 0 | -0.3 |
| Virginia | 88.0 ( 2.7) | 50.0 ( 1.5) | 6629083.2 ( 2.6) | -0.1 | 1.1 |
| Washington | 65.0 ( 2.0) | 25.9 ( 0.8) | 6023155.8 ( 2.4) | 0.4 | 1.6 |
| West Virginia | 21.0 ( 0.6) | 45.4 ( 1.4) | 1418397.0 ( 0.6) | 0 | -0.8 |
| Wisconsin | 62.0 ( 1.9) | 41.5 ( 1.3) | 4594462.8 ( 1.8) | -0.1 | 0.5 |
| Wyoming | 13.0 ( 0.4) | 45.4 ( 1.4) | 448159.5 ( 0.2) | -0.2 | -1.2 |
| Metro county | 2816.0 (85.3) | 2758.6 (83.5) | 222493432.8 (87.3) | 2 | 3.8 |
| Household Size | | | | | |
| 1 | 698.0 (21.1) | 609.6 (18.5) | 47084861.9 (18.5) | -2.6 | 0 |
| 2 | 1155.0 (35.0) | 1184.5 (35.9) | 86861158.7 (34.1) | -0.9 | -1.8 |
| 3 | 508.0 (15.4) | 554.8 (16.8) | 44465083.9 (17.5) | 2.1 | 0.7 |
| 4 | 609.0 (18.4) | 507.9 (15.4) | 41341780.4 (16.2) | -2.2 | 0.8 |
| 5 | 246.0 ( 7.5) | 263.8 ( 8.0) | 21311897.3 ( 8.4) | 0.9 | 0.4 |
| more than 5 | 86.0 ( 2.6) | 181.3 ( 5.5) | 13671405.1 ( 5.4) | 2.8 | -0.1 |
| Family Income | | | | | |
| Less than $10,000 | 205.0 ( 6.3) | 307.5 ( 9.5) | 9980755.3 ( 3.9) | -2.4 | -5.6 |
| $10,000 to $19,999 | 279.0 ( 8.6) | 312.3 ( 9.7) | 17761201.2 ( 7.0) | -1.6 | -2.7 |
| $20,000 to $29,999 | 298.0 ( 9.1) | 334.6 (10.3) | 21676442.2 ( 8.5) | -0.6 | -1.8 |
| $30,000 to $39,999 | 276.0 ( 8.5) | 301.4 ( 9.3) | 25273197.6 ( 9.9) | 1.4 | 0.6 |
| $40,000 to $49,999 | 243.0 ( 7.5) | 288.6 ( 8.9) | 19051200.4 ( 7.5) | 0 | -1.4 |
| $50,000 to $59,999 | 250.0 ( 7.7) | 292.5 ( 9.0) | 20844810.6 ( 8.2) | 0.5 | -0.8 |
| $60,000 to $99,999 | 786.0 (24.1) | 770.7 (23.8) | 60155849.0 (23.6) | -0.5 | -0.2 |
| $100,000 to $149,999 | 583.0 (17.9) | 400.3 (12.4) | 38001565.2 (14.9) | -3 | 2.5 |
| $150,000 or more | 339.0 (10.4) | 226.8 ( 7.0) | 41991165.9 (16.5) | 6.1 | 9.5 |
| Marital Status | | | | | |
| Married | 1989.0 (60.8) | 1794.1 (55.0) | 130042853.3 (51.1) | -9.7 | -3.9 |
| Widowed | 212.0 ( 6.5) | 201.0 ( 6.2) | 14914800.7 ( 5.9) | -0.6 | -0.3 |
| Divorced | 407.0 (12.4) | 383.8 (11.8) | 25557773.8 (10.0) | -2.4 | -1.8 |
| Separated | 54.0 ( 1.6) | 62.3 ( 1.9) | 4363383.0 ( 1.7) | 0.1 | -0.2 |
| Never married | 611.0 (18.7) | 822.5 (25.2) | 79857376.4 (31.3) | 12.6 | 6.1 |

Note. Employing post-stratification weights significantly reduces the differences of survey estimates from the target estimates including marital status that was not considered in raking procedures.

**Table S7. Summary of network characteristics before and after survey raking.**

| | unweighted | weighted | Diff. in percentage |
|---|---|---|---|
| N | 41033 | 41034.8 | |
| Network Size | | | |
| 0 | 4499.0 (11.0) | 5575.5 (13.6) | -2.6 |
| 1 | 16593.0 (40.4) | 17002.7 (41.4) | -1 |
| 2 | 10507.0 (25.6) | 9893.1 (24.1) | 1.5 |
| 3 | 5267.0 (12.8) | 4775.9 (11.6) | 1.2 |
| 4 | 2308.0 ( 5.6) | 2126.5 ( 5.2) | 0.4 |
| 5 | 865.0 ( 2.1) | 767.4 ( 1.9) | 0.2 |
| 6 | 994.0 ( 2.4) | 893.7 ( 2.2) | 0.2 |
| Relationship Composition (%) | | | |
| Parent | 14 | 12.1 | 1.9 |
| Spouse | 29.5 | 32.6 | -3.1 |
| Child | 8.8 | 9.6 | -0.8 |
| Sibling | 8.9 | 8.7 | 0.2 |
| Other family | 6.1 | 6.4 | -0.3 |
| Coworker | 0 | 0 | 0 |
| Friend | 23.8 | 21.6 | 2.2 |
| Neighbor | 1.4 | 1.4 | 0 |
| Other | 4.3 | 4.3 | 0 |
| Communication channel (%) | | | |
| In-person | 61.7 | 63.9 | -2.2 |
| Phone | 44.4 | 43.3 | 1.1 |
| Video | 16.2 | 13.4 | 2.8 |
| Text message | 35.8 | 32.4 | 3.4 |
| Email | 8.3 | 7.4 | 0.9 |
| SNS | 10.4 | 8.5 | 1.9 |
| Other channel | 1.8 | 1.8 | 0 |
| Homophily (i.e., % Same Alters) | | | |
| Sex | 51.8 | 48.2 | 3.6 |
| Race | 85.5 | 83 | 2.5 |
| Age | 57.4 | 58.2 | -0.8 |
| Education | 53.5 | 51.6 | 1.9 |
| Partisanship | 59.6 | 55.7 | 3.9 |